\documentclass[a4paper,12pt]{article}
\usepackage[top=25truemm,bottom=25truemm,left=15truemm,right=15truemm]{geometry}
\usepackage{hyperref}
\usepackage{graphicx} 
\usepackage{amsmath}
\usepackage{amssymb}
\graphicspath{{./figs/}}

\title{Complexity growth of rotating black holes with a probe string}
\author{Koichi Nagasaki}
\date{\today}

\begin{document}
\vspace{1cm}
\begin{center}
{\LARGE Complexity growth of rotating black holes with a probe string}\\
\vspace{2cm}
{\large Koichi Nagasaki}\footnote{koichi.nagasaki24@gmail.com}\\
\vspace{1cm}
{\small School of Physics,
University of Electronic Science and Technology of China (UESTC)\\
Address: No.4, Section 2, North Jianshe Road, Chengdu 610054, China}
\end{center}
\vspace{2cm}
\abstract{We study the effect of a probe string to black hole complexity according to the CA (Complexity equals Action) conjecture.
Our system contains a particle moving on the boundary of black hole spacetime.
In the dual description this corresponds to the insertion of a fundamental string on the bulk spacetime.
The total action consists of the Einstein-Hilbert term and the Nambu-Goto term. 
The effect of this string is expressed by the Nambu-Goto term.
Focusing on the Nambu-Goto term, we analyse the time development of this system.
Our results show some interesting properties of complexity.
This gives a useful hint for defining complexity in quantum field theories.}
\vspace{1cm}

\tableofcontents

\section{Introduction}
A concept of computational complexity is originally known in quantum information theory 
\cite{0034-4885-75-2-022001, TCS066, Dvali:2016lnb, Swingle:2016var, 
Hashimoto:2017fga, 2008arXiv0804.3401W, Bao:2018ira} 
or computer science 
\cite{Arora:2009:CCM:1540612, Moore:2011:NC:2086753}.
This concept gave a new physical quantity to study a gravitational physics.
An important goal of quantum gravity is to reveal the inside of the black hole horizon or information problem of black holes \cite{Coleman:1991ku, Preskill:1992tc, Giddings:1993vj, Russo:2005aw, Sekino:2008he, Terno:2009cc, Hartman:2013qma, Bradler:2013gqa, Polchinski:2016hrw, Marolf:2017jkr}. 
A candidate of its solution is firewalls \cite{Susskind:2012rm, Almheiri:2012rt, Stoltenberg:2014pua}.
Complexity is realized as an important quantity for studying the structure of black hole spacetime --- the Einstein-Rosen bridge \cite{Hayden:2007cs, Harlow:2013tf, Susskind:2014moa, Mann:2015luq, Barbon:2015ria, Barbon:2015soa, Couch:2016exn, Fu:2016xaa, Zhao:2017iul}, which is a structure of connecting two external black holes thought to be equivalent to entangled pair of particles (ER = EPR) \cite{Maldacena:2013xja}.
Especially, complexity is thought to be a good tool of diagnosing the existence of firewalls \cite{Zhao:2017iul}. 
Because of these motivations, complexity is studied in many recent works \cite{Susskind:2014moa, Stanford:2014jda, Susskind:2014rva, Susskind:2015toa, Roberts:2016hpo, Brown:2016wib, Cottrell:2017ayj, Reynolds:2017jfs, Yang:2017nfn, Zangeneh:2017tub, Susskind:2018fmx, Khan:2018rzm}.

The definite approach for quantifying complexity is still unknown.
So to define complexity in quantum field theory is one of theme of recent researches \cite{Vanchurin:2016met, Chapman:2017rqy, Jiang:2018gft, Molina-Vilaplana:2018sfn, Bhattacharyya:2018wym, Reynolds:2018zll}.
In the perspective of quantum information theory, complexity is roughly defined as the number of necessary gates which operate to produce the target state from the reference (initial) state.
The tensor network is frequently used in quantum system 
\cite{PhysRevLett115180405, May:2016dgv, Bhattacharyya:2016hbx, Bao:2017qmt, May:2017vyo, Caputa:2017yrh} 
and it is used for describing the warmhole structure \cite{Peach:2017npp}.
Then it seems to be a good approach for defining complexity.
There is also geometric approach for defining complexity \cite{Susskind:2014jwa}.
Geometric approach is suggested to introduce Finsler geometry on quantum space \cite{2005quant.ph2070N, 2006Sci311.1133N, 2007quant.ph1004D, Jefferson:2017sdb, Yang:2018nda, Hackl:2018ptj}.
In this approach complexity is determined by the geodesic on the quantum space.

The holographic duality \cite{Maldacena:1997re} is an important principle in recent researches.
According to this principle, complexity is expected to have a holographic dual in the gravity theory.
That relation between a gravity theory and a quantum field theory is recent active research theme \cite{Alishahiha:2015rta, Chemissany:2016qqq, Ge:2017rak, Czech:2017ryf, HosseiniMansoori:2017tsm, Moosa:2017yiz, Auzzi:2018zdu}.
The Complexity-Action (CA) conjecture \cite{Brown:2015bva, Brown:2015lvg} is the most reliable candidate for this duality. 
This conjecture suggest a relation between computational complexity and the gravitational action which is evaluated in a specific region of spacetime called a Wheeler-DeWitt (WDW) patch.
CA conjecture is tested in various spacetime setting \cite{Pan:2016ecg, Momeni:2016ekm, Chapman:2016hwi, Lehner:2016vdi, Carmi:2016wjl, Tao:2017fsy, Alishahiha:2017hwg, Reynolds:2017lwq, Qaemmaqami:2017lzs, Guo:2017rul, Miao:2017quj, Sebastiani:2017rxr, Couch:2017yil, Swingle:2017zcd, Cano:2018aqi, Ghaffarnejad:2018bsd, Chapman:2018dem, Chapman:2018lsv, Fareghbal:2018ngr, Auzzi:2018pbc, Ghaffarnejad:2018prc, Alishahiha:2018tep, An:2018xhv}.
Complexity growth of some kinds of black holes, especially Kerr-AdS, are calculated in \cite{Cai:2016xho}.
In general there is the divergence of the action.
Treating the boundary terms of the gravitational action and its renormalization is one of the important problem for proving this conjecture.
For this purpose the Neumann boundary term for gravity \cite{Krishnan:2016mcj} and other solutions are considered so far \cite{Parattu:2015gga, Reynolds:2016rvl, Carmi:2016wjl, Chakraborty:2016yna, Kim:2017lrw, Gan:2017qkz, Chakraborty:2017zep,
Chakraborty:2018dvi, Jiang:2018sqj}.
That conjecture in the time dependent system \cite{Pan:2016ecg, Momeni:2016ira, Huang:2016fks, Carmi:2017jqz, Kim:2017qrq, Ghodrati:2017roz, Jiang:2018pfk} is our main interest here.
In \cite{Kim:2017qrq}, especially, the counterpart in a field theory is discussed for Finsler geometry and Fubini-Study metric.

Some of the property of complexity is found in recent works.
For example, it has a good analogy with entropy in thermodynamics:
it satisfies the second law of thermodynamics \cite{Brown:2017jil}.
The time development of complexity satisfies the Lloyd bound \cite{Brown:2015lvg, 2000Natur4061047L, Moosa:2017yvt}.
Especially, in \cite{Moosa:2017yvt} complexity in the process of the formation of the black holes is discussed.
Furthermore, interestingly, \cite{Fu:2018kcp} revealed that complexity has a nonlocal property.

The analysis of complexity using a probe is a useful method. 
For example, in \cite{Abad:2017cgl} complexity growth in a system with flavor branes is studied and a nonlocal operator in BTZ black hole is studied in \cite{Ageev:2014nva}.
And also complexity of particle falling in the Poincare-AdS in the probe approximation in \cite{Ageev:2018nye}.
In this paper we use these kind of method. 
Our probe is a fundamental string. 
The Einstein-Hilbert action of various kinds of spacetime is calculated in many works.
On the other hand we study the effect of the probe string here.
The total action of this system is the sum of the Einstein-Hilbert action and the Nambu-Goto (NG) action of the string. 
The NG action is also obtained by integrating over the WDW patch.
The motivation for studying such effect is found in \cite{Gubser:2006bz, Herzog:2006gh, CasalderreySolana:2006rq, Liu:2006ug, Fadafan:2008bq, Fadafan:2012qu, Atashi:2016cso} where energy loss of the charged quarks are calculated by considering a drag force caused by the string motion.

In the previous paper \cite{Nagasaki:2017kqe} we studied the effect of the probe string moving on the AdS$_{3+1}$ black hole spacetime.
What I found about complexity so far is:
\begin{itemize}
\item Complexity basically grows as the black hole mass becomes larger.
\item But in the vicinity of the light speed complexity shows a specific behavior.
\item Complexity is smaller as the probe string moves faster.
\end{itemize}
The most notable result is the last one.
I think it can be stated that a fast moving object decreases the growth of the complexity.
These result may serve a hint to find the definition of complexity in quantum field theory.
In this paper we tried to find new properties of complexity by studying the effects of probe string on more broad type of black hole. 

This paper is organized as follows:
In section \ref{sec:AdSBH} we begin with calculating the effect of the probe string in the AdS black hole. 
This is the higher dimensional generalization of the previous work.
We first compute the general $n$-dimensional case and reproduce the $(4+1)$-dimensional results.
And then $(3+1)$ and $(5+1)$-dimensional results are also found.
In section \ref{sec:BTZBH} we study the NG action of a string moving in three dimensional black hole spacetime.
In this section a new parameter --- angular momentum is introduced.
This black hole is the BTZ black hole \cite{Banados:1992wn}.
The angular momentum will show an interesting phenomena which is not found in our previous work.
In section \ref{sec:KABH} the angular momentum is added to the AdS black holes. 
This is the Kerr-AdS black hole.
Their complexity growth is studied in \cite{Cai:2016xho}.
The drag force of the four-dimensional Kerr-AdS black hole is studied in \cite{NataAtmaja:2010hd}.
In their case, the drag force locates in the boundary of the AdS.
In our case, on the other hand, we take into account the inner of the black hole horizon.
We review this analysis and also study the five dimensional case.
In the final section \ref{sec:Discussion} we summarize our results and remark the new insight about them.
After that some future directions are suggested.

\section{AdS$_{n+1}$ black holes}\label{sec:AdSBH}
In this section we study the cases of AdS black hole in arbitrary dimension.
The $n=3$ case will reproduce the previous work \cite{Nagasaki:2017kqe}.
Here we consider non charged black holes.
This metric is
\begin{align}
ds_\text{AdS$_{n+1}$}^2
&= -f(r)dt^2 + \frac{dr^2}{f(r)} + r^2d\Omega_{n-1},\label{eq:AdSmetric(n+1)}\\
f(r) 
&= 1 - \frac{8\pi}{(n-1)\Omega_{n-1}}\frac{2GM}{r^{n-2}}
  + \frac{r^2}{\ell_\text{AdS}^2}
= 1 - \frac{r_\text{m}^{n-2}}{r^{n-2}} + \frac{r^2}{\ell_\text{AdS}^2},\qquad
r_\text{m}^{n-2} := \frac{16\pi GM}{(n-1)\Omega_{n-1}}.\label{eq:AdSmetricfunc(n+1)}
\end{align}
The volume of $(n-1)$-sphere is $\Omega_{n-1} = 2\pi^{n/2}/\Gamma(n/2)$.
In the each dimension the relation between $r_\text{m}$ and mass is from
$r_\text{m}^{n-2} = 16\pi GM/((n-1)\Omega_{n-1})$.
For later use, we write them here explicitly in four, five and six dimensions:
\begin{subequations}\label{eq:relrm&M}
\begin{align}
\text{(3+1)-dim: }&
r_\text{m}
= \frac{8\pi M}{\Omega_2}
= 2M,\\
\text{(4+1)-dim: }&
r_\text{m}
= \Big(\frac{16\pi M}{3\Omega_3}\Big)^{1/2}
= \Big(\frac{8M}{3\pi}\Big)^{1/2},\\
\text{(5+1)-dim: }&
r_\text{m}
= \Big(\frac{4\pi M}{\Omega_4}\Big)^{1/3}
= \Big(\frac{3M}{2\pi}\Big)^{1/3}.
\end{align}
\end{subequations} 

As before we assume that the string moves a great circle on $S^{n-1}$ subspace.
Then the induced metric of this part is as the same as bofore $d\Omega_{n-1} = d\phi^2$.
We take the worldsheet parameter as \eqref{eq:WScoord}:
\begin{equation*}
t = \tau,\; r = \sigma,\; \phi = v\tau + \xi(\sigma).
\end{equation*}
As before we scale $r$ so that $\ell_\text{AdS} = 1$.
In the following $t$, $r_\text{m}$, $M$ and the worldsheet coordinate $\sigma$ are rescaled in the same way.
Then $\ell_\text{AdS}$ disappears in the expression and the metric is rescaled the original one times $\ell_\text{AdS}^2$.
The induced metric is 
\begin{align}\label{eq:dsAdS(n+1)ind}
ds^2_\text{AdS$_{n+1}$ind} 
&= (-f(\sigma) + \sigma^2v^2)d\tau^2 
  + \Big(\frac{1}{f(\sigma)} + \sigma^2\xi'(\sigma)^2\Big)d\sigma^2 
  + 2\sigma^2v\xi'(\sigma)d\tau d\sigma,\\
f(\sigma) 
&= 1 - \Big(\frac{r_\text{m}}{\sigma}\Big)^{n-2} + \sigma^2.\nonumber
\end{align}
The NG action is obtained by integrating over the WDW patch, 
\begin{equation}
\frac{dS_\text{NG}}{dt}
= T_\text{s}\int_0^{r_\text{h}} d\sigma\sqrt{-g_\text{ind}(\sigma)} 
= T_\text{s}\int_0^{r_\text{h}} d\sigma
   \sqrt{1 - \frac{v^2\sigma^2}{f(\sigma)} + \sigma^2f(\sigma)\xi'(\sigma)^2}
=: \int_0^{r_\text{h}} d\sigma \mathcal L_\text{AdS(n+1)},
\end{equation}
where the horizon $r_\text{h}$ is determined by $f(r)=0$.
Here we commet on the horizon.
By differentiating $f(r)$,
\begin{equation}
f'(r) = (n-2)\frac{r_\text{m}}{r^{n-1}}+2r; \; n\geq 3,
\end{equation}
we find that $f(r)$ is a monotonically increasing function.
This fact and this function takes a negative value near $r=0$ mean the equation $f(r)=0$ certainly has only one positive solution.

\paragraph{EOM and its solution}
The equation of motion for $\xi$ gives
\begin{equation}
0 = \frac{d}{d\sigma}\Big(
  \frac{\sigma^2f(\sigma)\xi'(\sigma)}{\sqrt{1-v^2\sigma^2/f(\sigma) + \sigma^2f(\sigma)\xi'(\sigma)^2}}\Big).
\end{equation}
From this equation the constant $c_\xi$ is defined as follows:
\begin{equation}
c_\xi := \frac{\sigma^2f(\sigma)\xi'(\sigma)}{\mathcal L_\text{AdS(n+1)}/T_\text{s}}.
\end{equation}
By solving it for $\xi'(\sigma)$, 
\begin{equation}
\xi'(\sigma)
= \frac{c_\xi}{\sigma^2f(\sigma)}
  \sqrt\frac{\sigma^2f(\sigma) - v^2\sigma^4}{\sigma^2f(\sigma)-c_\xi^2}.\label{eq:solxindim}
\end{equation}
The constant $c_\xi$ is determined in the same way as before from the real valued condition.
The zero of the numerator gives the equation:
\begin{equation}\label{eq:realcondnum}
f(\sigma) - v^2\sigma^2 = 0
\Rightarrow
(1-v^2)\sigma^n + \sigma^{n-2} - r_\text{m}^{n-2} = 0.
\end{equation}
The function in the left hand side is a monotonically increasing function of $\sigma$ (assumed that $n\geq 3$) and takes a negative value at $\sigma=0$.
Then this function has only one solution in $\sigma > 0$.
We call it $\sigma_\text{H}$:
$(1-v^2)\sigma_\text{H}^n + \sigma_\text{H}^{n-2} - r_\text{m}^{n-2} = 0$.
Since the denominator becomes zero at the same value of $\sigma$, $\sigma=\sigma_\text{H}$, the constant $c_\xi$ is determined:
\begin{equation}
0 = \sigma_\text{H}^2f(\sigma_\text{H}) - c_\xi^2
= v^2\sigma_\text{H}^4 - c_\xi^2,\;
\therefore
c_\xi = v\sigma_\text{H}^2.
\end{equation}
In the above the second equality is derived from numerator condition \eqref{eq:realcondnum}.
We assumed that $c_\xi$ is positive.
We obtain
\begin{equation}
\xi'(\sigma)
= \frac{c_\xi}{\sigma^2f(\sigma)}
  \sqrt\frac{\sigma^{n-2}f(\sigma) - v^2\sigma^n}{\sigma^{n-2}f(\sigma)-v^2\sigma_\text{H}^4\sigma^{n-4}}
\end{equation}

\paragraph{Action}
The NG action is obtained by integrating over the WDW patch:
\begin{equation}\label{eq:AdSNGaction(n+1)}
\frac{1}{T_\text{s}}\int_0^{r_\text{h}}d\sigma\mathcal L_\text{AdS(n+1)}
= \int_0^{r_\text{h}}d\sigma
	\sqrt\frac{\sigma^{n-2}f(\sigma) - v^2\sigma^n}{\sigma^{n-2}f(\sigma)-v^2\sigma_\text{H}^4\sigma^{n-4}}. 
\end{equation}
This form is general form for $n\geq 3$. 
In the following we focus concretely on four, five and six dimensions.

\subsection{AdS$_{3+1}$ case}
In $(3+1)$-dimension, eq.\eqref{eq:AdSNGaction(n+1)} is 
\begin{equation}
\xi'(\sigma)
= \frac{c_\xi}{\sigma^2f(\sigma)}
  \sqrt\frac{\sigma f(\sigma) - v^2\sigma^3}{\sigma f(\sigma)-v^2\sigma_\text{H}^4/\sigma},\;
 f(\sigma) = 1 - \frac{r_\text{m}}{\sigma} + \sigma^2.\label{eq:fAdS(3+1)}
\end{equation}
By construction the numerator and the denominator have the common factor. 
Then the expression can be simplified.
\begin{align*}
\sigma f(\sigma) - v^2\sigma^3
- (\sigma_\text{H}f(\sigma_\text{H}) - v^2\sigma_\text{H}^3)
&= (\sigma-\sigma_\text{H})
 (1 + (1-v^2)(\sigma^2 + \sigma_\text{H}\sigma + \sigma_\text{H}^2)),\\
\sigma f(\sigma) - v^2\sigma_\text{H}^4/\sigma
- (\sigma_\text{H}f(\sigma_\text{H}) - v^2\sigma_\text{H}^3)
&= (\sigma-\sigma_\text{H})
 (1 + (\sigma^2 + \sigma_\text{H}\sigma + \sigma_\text{H}^2)
     + v^2\sigma_\text{H}^4/(\sigma\sigma_\text{H})).
\end{align*}
Then the development of the NG action is 
\begin{align}
\frac{1}{T_\text{s}}\frac{dS_\text{NG}}{dt}
&= \int_0^{r_\text{h}}d\sigma
\sqrt\frac{1 + (1-v^2)(\sigma^2 + \sigma\sigma_\text{H} + \sigma_\text{H}^2)}
  {1 + (\sigma^2 + \sigma\sigma_\text{H} + \sigma_\text{H}^2) + v^2\sigma_\text{H}^3/\sigma}\nonumber\\
&= \int_0^{r_\text{h}}d\sigma
\sqrt\frac{(1-v^2)(\sigma^3 + \sigma_\text{H}\sigma^2) + ((1-v^2)\sigma_\text{H}^2+1)\sigma}
  {\sigma^3 + \sigma_\text{H}\sigma^2 + (\sigma_\text{H}^2+1)\sigma +  v^2\sigma_\text{H}^3}.
\end{align}
By numerical calculation, this action can be expressed as a function of $M$ and $v$.
Recall that the black hole mass is given by eq.\eqref{eq:fAdS(3+1)} and eq.\eqref{eq:relrm&M}.
The result is shown in figures \ref{fig:AdS31ActionVelocity} and \ref{fig:AdS31ActionMass}.

The left one \ref{fig:AdS31ActionVelocity} shows the velocity dependence.
As usual this dependence takes a maximum when the string is stationary. 

The right one \ref{fig:AdS31ActionMass} shows the mass dependence.
There are notable behaviors here.
One is a peak at lower mass.
The other is a phase transition.
For slower strings, its effect increases according to mass increasing.
But fast strings, especially, near light speed, it changes to a decreasing function of the mass.

\subsection{AdS$_{4+1}$ case}
When $n=4$, \eqref{eq:AdSNGaction(n+1)} becomes
\begin{align}
\frac{1}{T_\text{s}}\frac{dS_\text{NG}}{dt}
= \int_0^{r_\text{h}}d\sigma
\sqrt\frac{(1-v^2)(\sigma^2+\sigma_\text{H}^2) + 1}{\sigma^2 + \sigma_\text{H}^2 + 1}.
\end{align}
As studied in \cite{Nagasaki:2017kqe}, this integral is performed by elliptic integral.
We show here the result again:
\begin{align}
\frac{dS_\text{NG}}{dt} 
&= -iT_\text{s}
 \Bigg(\frac{1+\sqrt{1+4r_\text{m}^2
    (1-v^2)}}{2}\Bigg)^{1/2}\times\nonumber\\
&\qquad
E\Bigg[\arcsin\Big(i\Big(
  \frac{(-1+\sqrt{4r_\text{m}^2+1})(1-v^2)}{(1 - 2v^2) +\sqrt{1+4r_\text{m}^2(1-v^2)}}\Big)^{1/2}\Big),
    \Big(\frac{(1 - 2v^2) + \sqrt{1+4r_\text{m}^2(1-v^2)}}  
  {1+\sqrt{1+4r_\text{m}^2(1-v^2)}}\Big)^{1/2}\:
  \Bigg],
\end{align}
where $r_m$ is related to the black hole mass by 
$4r_\text{m}^2 = 32M/(3\pi)$ as noted in \eqref{eq:relrm&M}.
The velocity dependence and the mass dependence are shown in figures \ref{fig:AdS41Actionvelocity} and \ref{fig:AdS41Actionmass}.
It reproduces the results in the previous work \cite{Nagasaki:2017kqe}.

\subsection{AdS$_{5+1}$ case}
\begin{figure}[h]
	\begin{minipage}[t]{0.5\linewidth}
	\includegraphics[width=\linewidth]{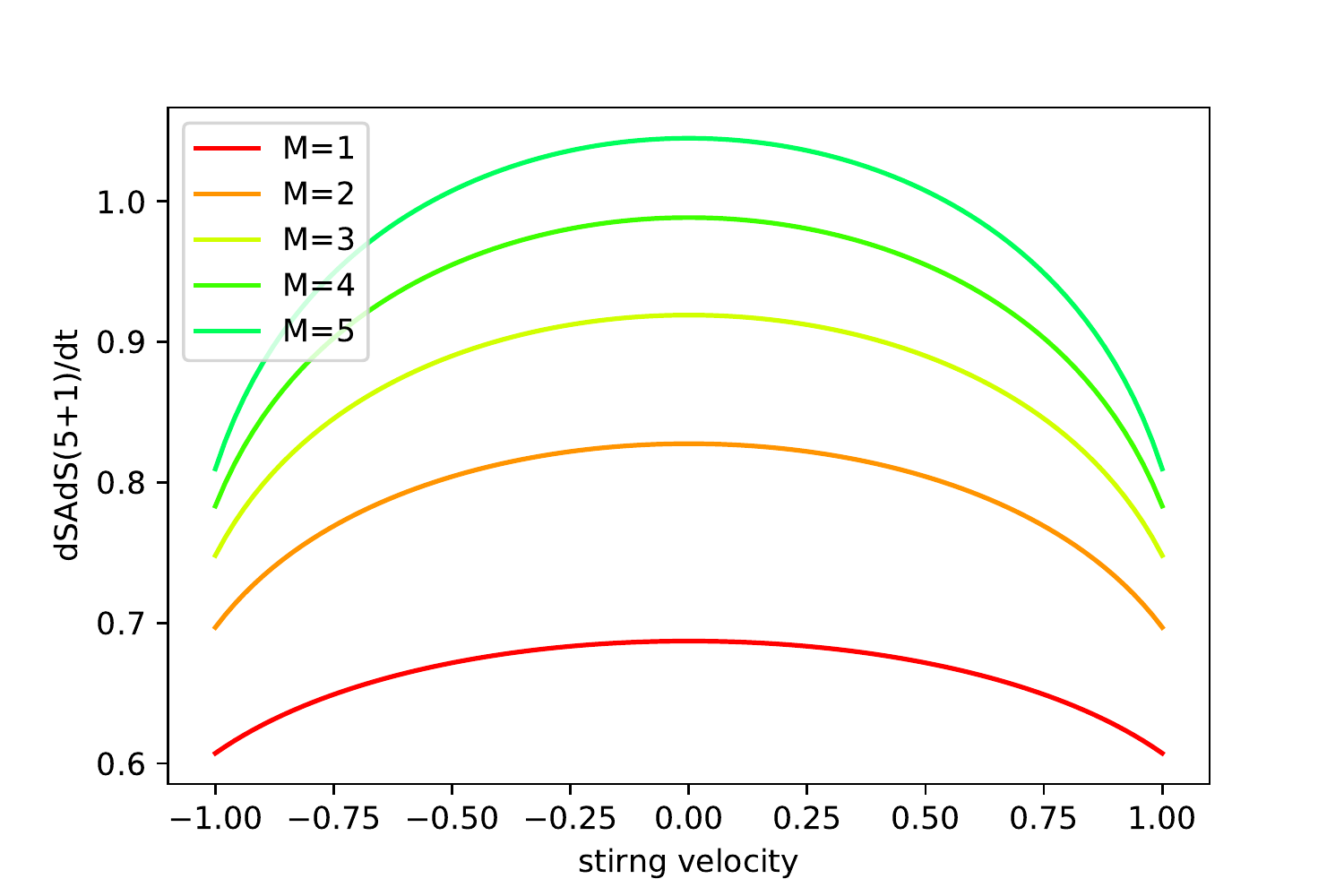}
	\caption{AdS$_{5+1}$: Action growth - string velocity}
	\label{fig:AdS51Actionvelocity}
	\end{minipage}
\hspace{0.01\linewidth}
	\begin{minipage}[t]{0.5\linewidth}
	\includegraphics[width=\linewidth]{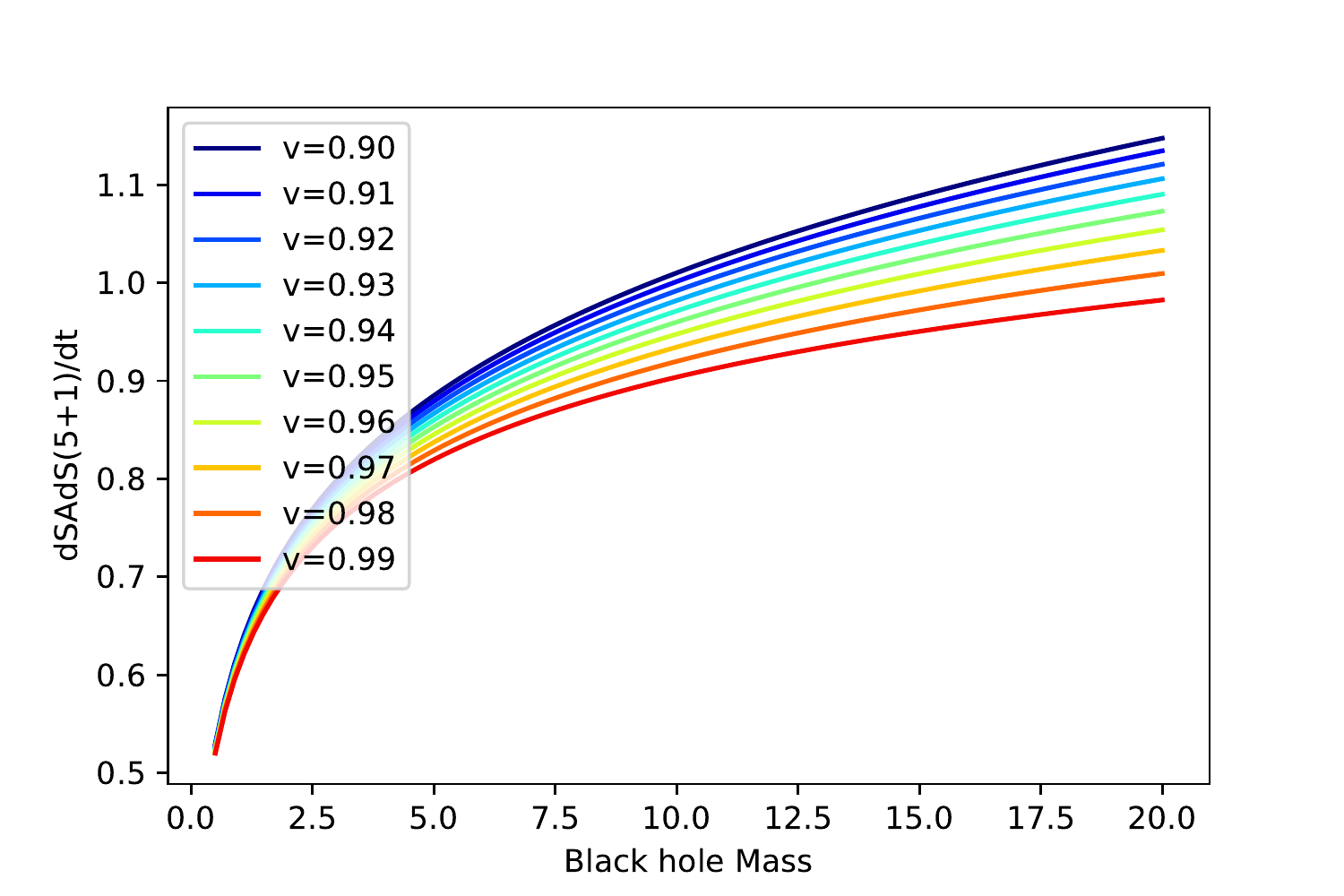}
	\caption{AdS$_{5+1}$: Action growth - Black hole mass}
	\label{fig:AdS51Actionmass}
	\end{minipage}
\end{figure}

In the (5+1)-dimensional case, eq.\eqref{eq:AdSNGaction(n+1)} is explicitly, 
\begin{align}
\frac{1}{T_\text{s}}\frac{dS_\text{NG}}{dt}
&= \int_0^{r_\text{h}}d\sigma
\sqrt\frac{(\sigma^2 + \sigma_\text{H}\sigma + \sigma_\text{H}^2)  
  + (1-v^2)(\sigma^4 + \sigma_\text{H}\sigma^3 + \sigma_\text{H}^2\sigma^2 + \sigma_\text{H}^3\sigma + \sigma_\text{H}^4)}
  {(\sigma^2 + \sigma_\text{H}\sigma + \sigma_\text{H}^2)
  + (\sigma^4 + \sigma_\text{H}\sigma^3 + \sigma_\text{H}^2\sigma^2 + \sigma_\text{H}^3\sigma + \sigma_\text{H}^4)  
  - v^2\sigma_\text{H}^4}.
\end{align}
This integral is also performed by the numerical calculation method.
The velocity dependence and the mass dependence are shown in the figures \ref{fig:AdS51Actionvelocity} and \ref{fig:AdS51Actionmass}.
Remarkable points of these plot are as follows.

\begin{figure}[h]
	\begin{minipage}[t]{0.5\linewidth}
	\includegraphics[width=\linewidth]{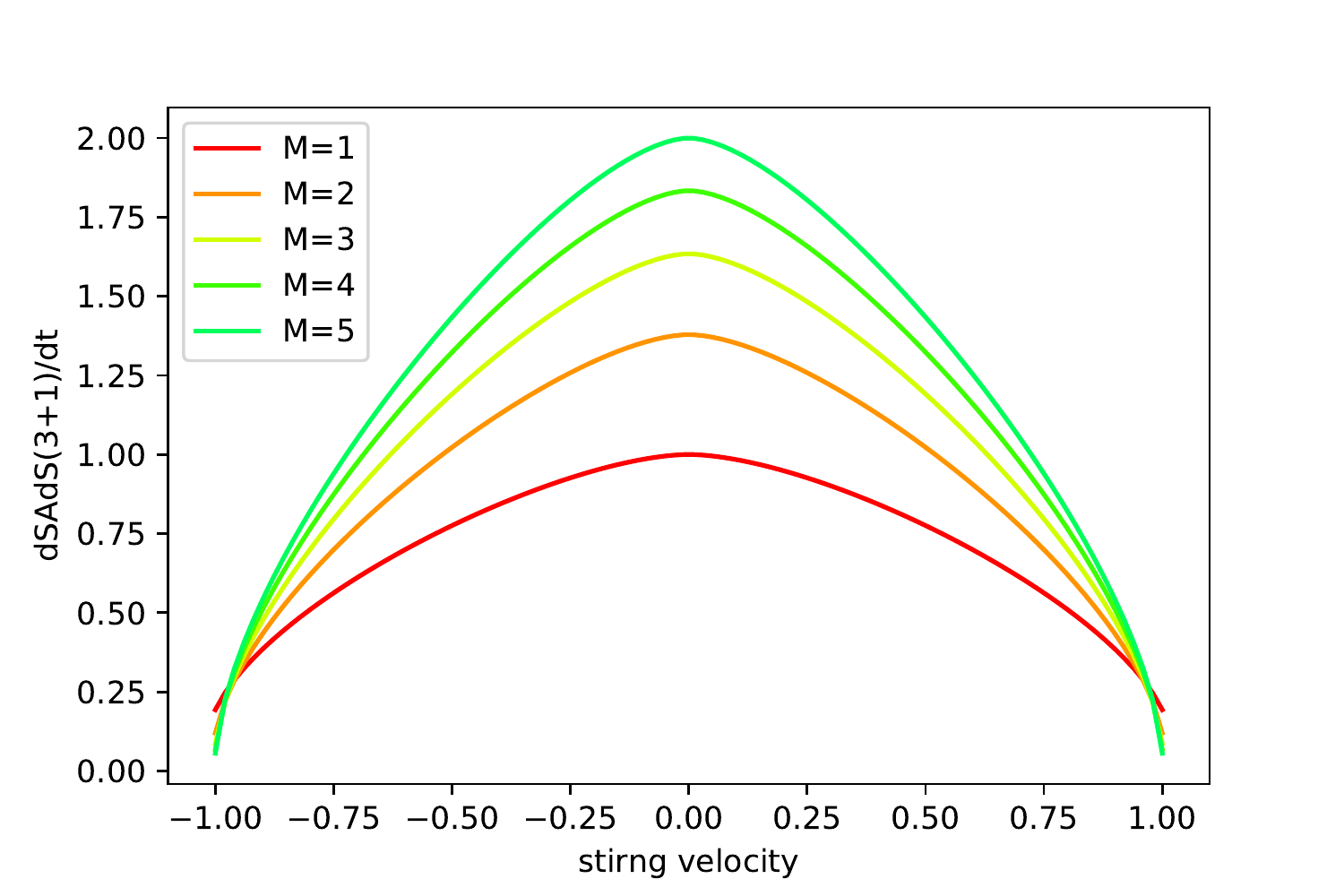}
	\caption{AdS$_{3+1}$: Action growth - string velocity}
	\label{fig:AdS31ActionVelocity}
	\end{minipage}
\hspace{0.01\linewidth}
	\begin{minipage}[t]{0.5\linewidth}
	\includegraphics[width=\linewidth]{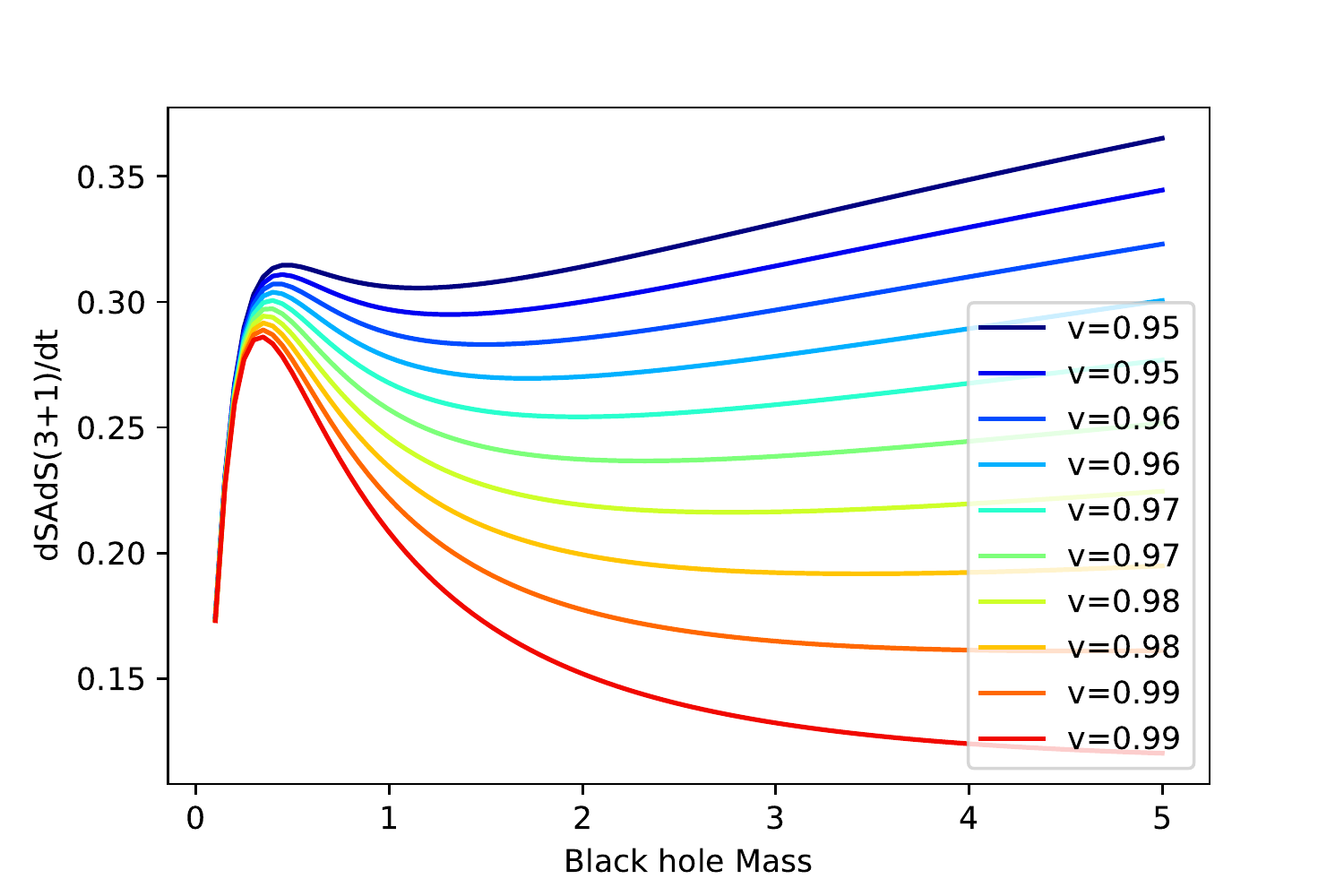}
	\caption{AdS$_{3+1}$: Action growth - Black hole mass}
	\label{fig:AdS31ActionMass}
	\end{minipage}
\end{figure}

\begin{figure}[h]
	\begin{minipage}[t]{0.5\linewidth}
	\includegraphics[width=\linewidth]{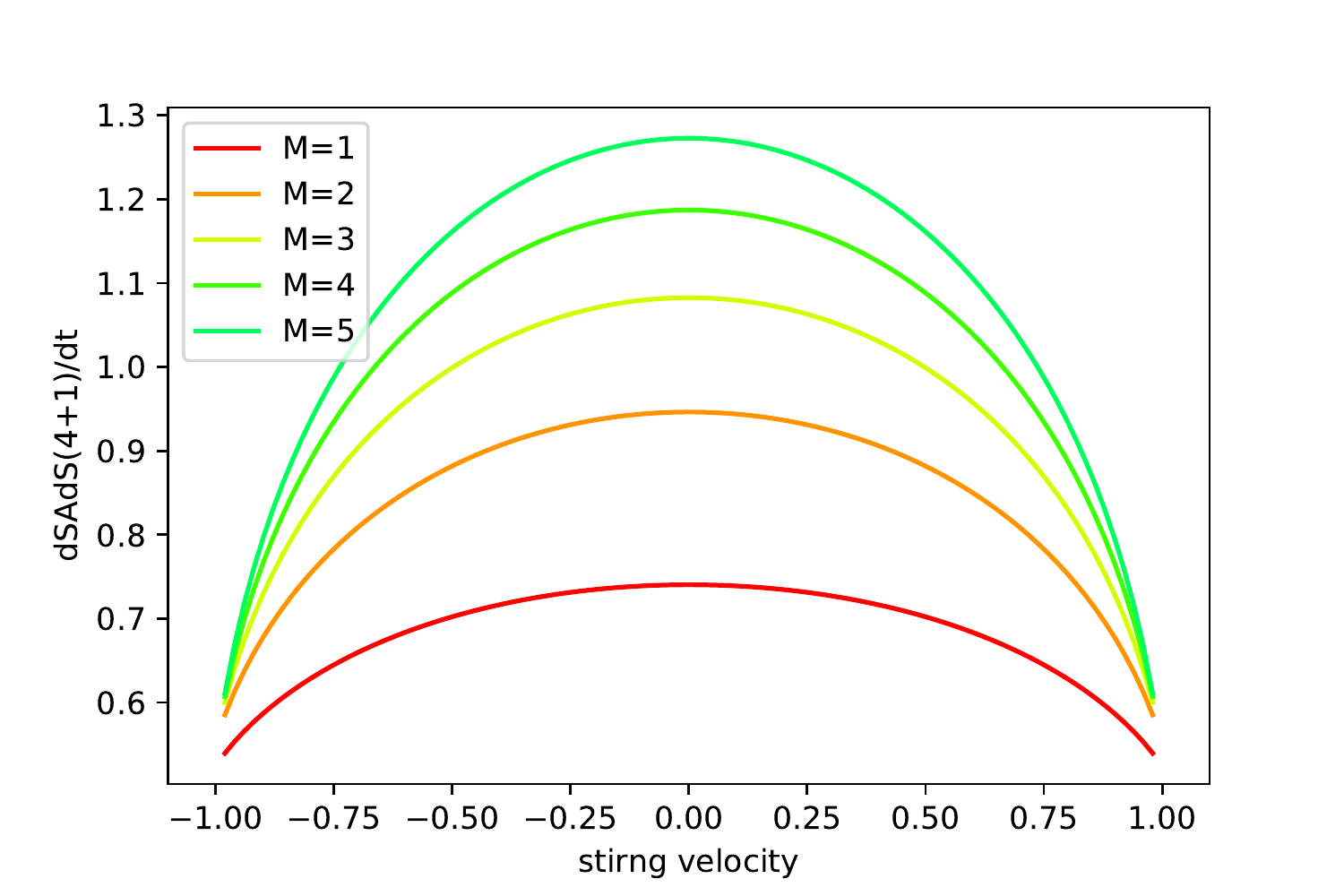}
	\caption{AdS$_{4+1}$: Action growth - string velocity}
	\label{fig:AdS41Actionvelocity}
	\end{minipage}
\hspace{0.01\linewidth}
	\begin{minipage}[t]{0.5\linewidth}
	\includegraphics[width=\linewidth]{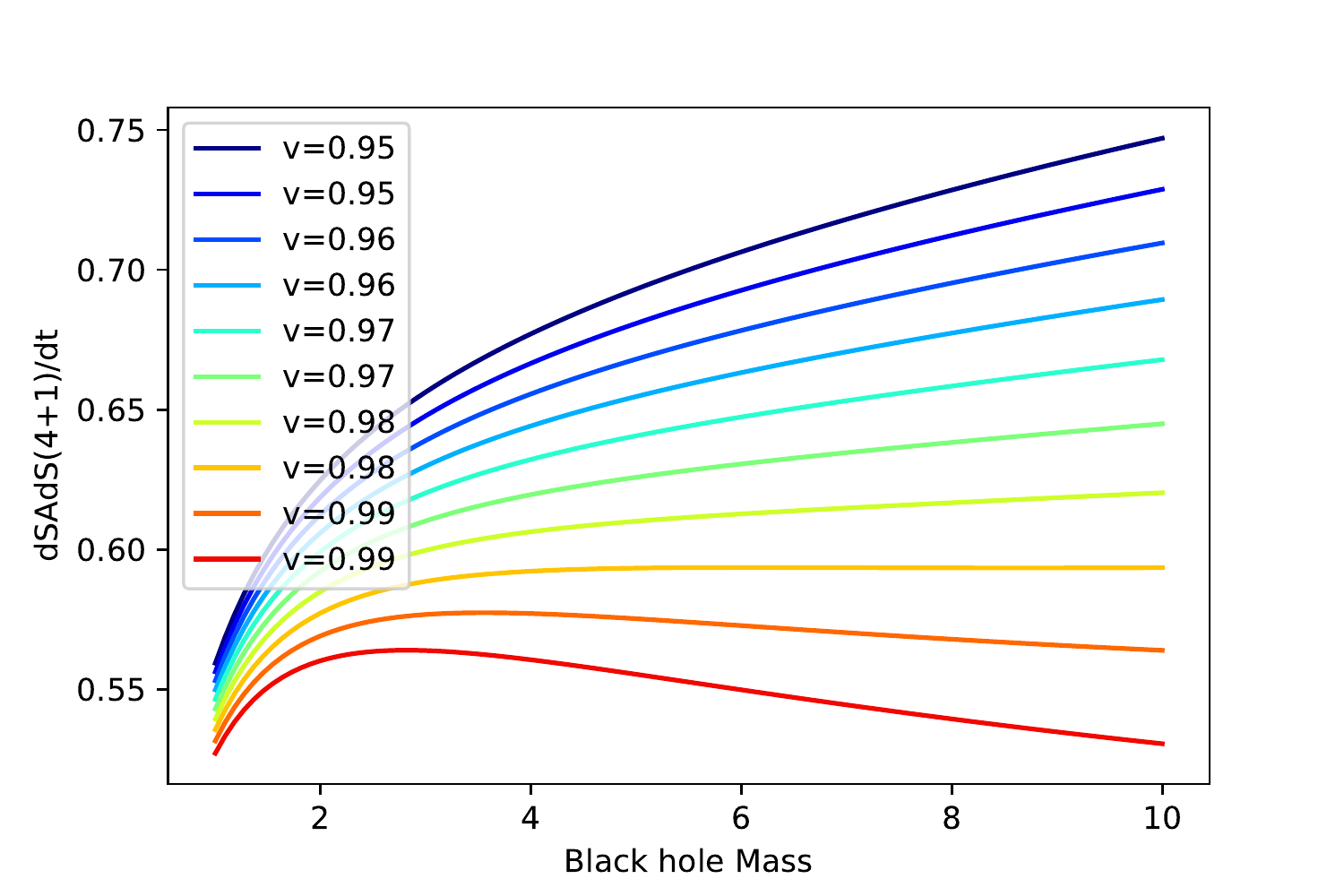}
	\caption{AdS$_{4+1}$: Action growth - Black hole mass}
	\label{fig:AdS41Actionmass}
	\end{minipage}
\end{figure}

For figure \ref{fig:AdS51Actionvelocity}, the curve of the velocity dependence is gentler as compared with four and five dimensional cases (\ref{fig:AdS31ActionVelocity} and \ref{fig:AdS41Actionvelocity}).
Compared with the lower dimensional cases, the effect of the probe string decreases slower in higher velocity.
Especially it does not reach to zero at the light velocity while it becomes zero in the BTZ black hole case (see figures \ref{fig:BTZActionvelocityJM09} and \ref{fig:BTZActionvelocityJM02}).
We can say that the effect to complexity becomes insensitive as the dimensionality is higher.
It can also be seen from the fact that the maximum value is lower than AdS$_{3+1}$ and AdS$_{4+1}$ cases. 

For the mass dependence \ref{fig:AdS51Actionmass} the maximum at the vicinity of the light speed disappears here.
This is now a monotonically increasing function of mass in all regions of mass and velocity.

We expect that these behaviors is a general tendency in higher than six dimension.
That is, the dependence between different masses and velocities become smaller in higher dimensions.

\section{BTZ black holes}\label{sec:BTZBH}
We consider the string moving in the BTZ black hole spacetime.
The butterfly effects caused by a small perturbation on an asymptotic region in this black holes is studied in \cite{Reynolds:2016pmi}.
In this section we study the effect of the string moving on this spacetime geometry. 
The BTZ black hole is $(2+1)$-spacetime specified as
\begin{equation}
ds_\text{BTZ}^2 
= -f(r)dt^2 + \frac{dr^2}{f(r)} 
 + r^2\Big(d\phi - \frac{r_{+}r_{-}}{\ell_\text{AdS}r^2}dt\Big)^2,\;
f(r) := \frac{(r^2-r_{+}^2)(r^2-r_{-}^2)}{\ell_\text{AdS}^2r^2}.
\end{equation}
The parameters $r_\pm$ are the inner and the outer horizon which is related to black hole mass $M$ and angular momentum $J$ by
\begin{equation}
M = (r_{+}^2 + r_{-}^2)/\ell_\text{AdS}^2,\;
J = 2r_{+}r_{-}/\ell_\text{AdS}.
\end{equation}
We rescale $r_\text{(old)} = r_\text{(new)}\ell_\text{AdS}$ and for $r_\pm$ and $t$ in the same way.
This simplifies the expression and the metric becomes the original one times $\ell_\text{AdS}^2$.
One edge of the string moves with velocity $v$.
We parametrize the worldsheet as follows:
\begin{equation}\label{eq:WScoord}
t = \tau,\; r = \sigma,\; \phi = v\tau + \xi(\sigma).
\end{equation}
Since we here use the $\ell_\text{AdS}=1$ unit, the angular velocity of the string $\omega$ is $\omega = v/\ell_{AdS} = v$. 
The induced metric is 
\begin{align}
ds_\text{BTZind}^2
&= -\Big(f(\sigma) - \Big(v\sigma- \frac{r_{+}r_{-}}{\sigma}\Big)^2\Big)d\tau^2 
  + \Big(\frac{1}{f(\sigma)} + \sigma^2\xi'(\sigma)^2\Big)d\sigma^2\nonumber\\
&\qquad
  + 2\Big(v\sigma- \frac{r_{+}r_{-}}{\sigma}\Big)
    \sigma\xi'(\sigma)d\tau d\sigma,\\
f(\sigma) &= \frac{(\sigma^2 - r_{+}^2)(\sigma^2 - r_{-}^2)}{\sigma^2}.
\end{align}
The Nambu-Goto Lagrangian is given by the determinant of this metric.
\begin{equation}
\mathcal L_\text{BTZ}
= T_\text{s}\sqrt{1 + f(\sigma)\sigma^2\xi'(\sigma)^2
  - \frac{1}{f(\sigma)}\Big(v\sigma - \frac{J}{2\sigma}\Big)^2},
\end{equation}
where angular momentum is rescaled as $J_\text{(old)}/\ell_\text{AdS} =: J$.

\paragraph{EOM and its solution}
By the equation of motion,
\begin{align}
&\frac{d}{d\sigma}\Big(\frac{f(\sigma)\sigma^2\xi'(\sigma)}{\mathcal L_\text{BTZ}/T_\text{s}}\Big) = 0,\;
c_\xi := \frac{f(\sigma)\sigma^2\xi'(\sigma)}{\mathcal L_\text{BTZ}/T_\text{s}},\label{eq:cxifsigmaBTZ}\\
&\xi'(\sigma) 
= \frac{c_\xi}{\sigma^2f(\sigma)}
  \sqrt\frac{\sigma^2f(\sigma)-(v\sigma^2-J/2)^2}{\sigma^2f(\sigma)-c_\xi^2}.\label{eq:xiprimeBTZ}
\end{align}
For this function to give the real values, the numerator and the denominator in the square root must have the same zero point.
This condition leads that the denominator is zero when 
\begin{equation}
\sigma 
= \sigma_\text{H}^2
:= \frac{M-vJ}{1-V^2}.\label{eq:zeropBTZ}
\end{equation}
This determines the integration constant as
\begin{equation}
c_\xi = |v\sigma_\text{H}^2-J/2|.
\end{equation}
Then the square root of \eqref{eq:xiprimeBTZ} is factorized by $(\sigma-\sigma_\text{H})$.
\begin{equation}
\xi'(\sigma)
= \frac{c_\xi}{\sigma^2f(\sigma)}
 \sqrt\frac{(1 - v^2)(\sigma^2 + \sigma_\text{H}^2) - (M-vJ)}{\sigma^2 + \sigma_\text{H}^2 - M}.
\end{equation}
From the relation \eqref{eq:cxifsigmaBTZ}, the Lagrangian is 
\begin{align}
\frac{\mathcal L_\text{BTZ}}{T_\text{s}} 
= \sqrt\frac{(1-v^2)\sigma^2}{\sigma^2 + \sigma_\text{H}^2 - M}.
\end{align}

\begin{figure}[h]
	\begin{minipage}[t]{0.5\linewidth}
	\includegraphics[width=\linewidth]{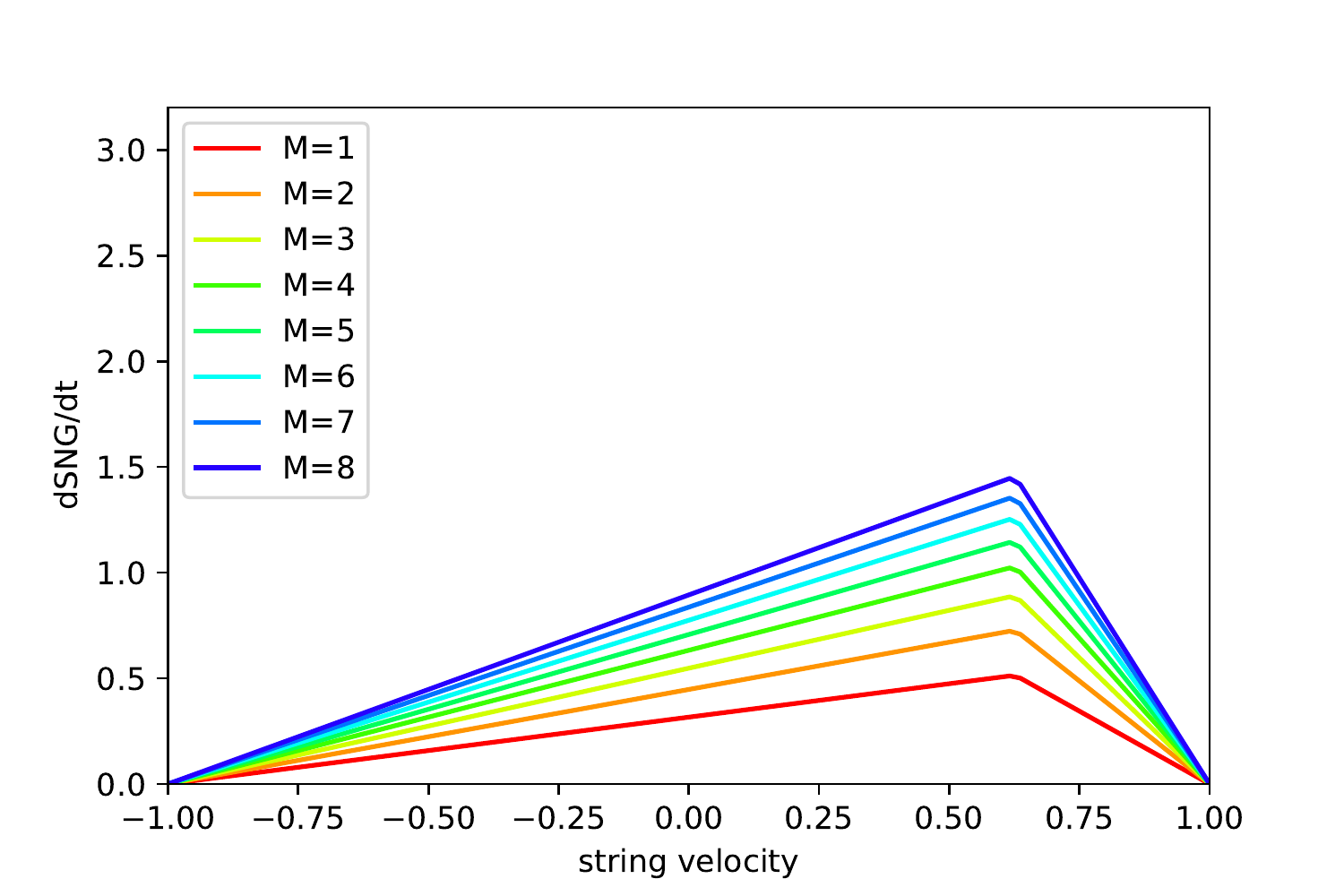}
	\caption{BTZ: Action growth - string velocity $J/M = 0.9$ fixed}
	\label{fig:BTZActionvelocityJM09}
	\end{minipage}
\hspace{0.01\linewidth}
	\begin{minipage}[t]{0.5\linewidth}
	\includegraphics[width=\linewidth]{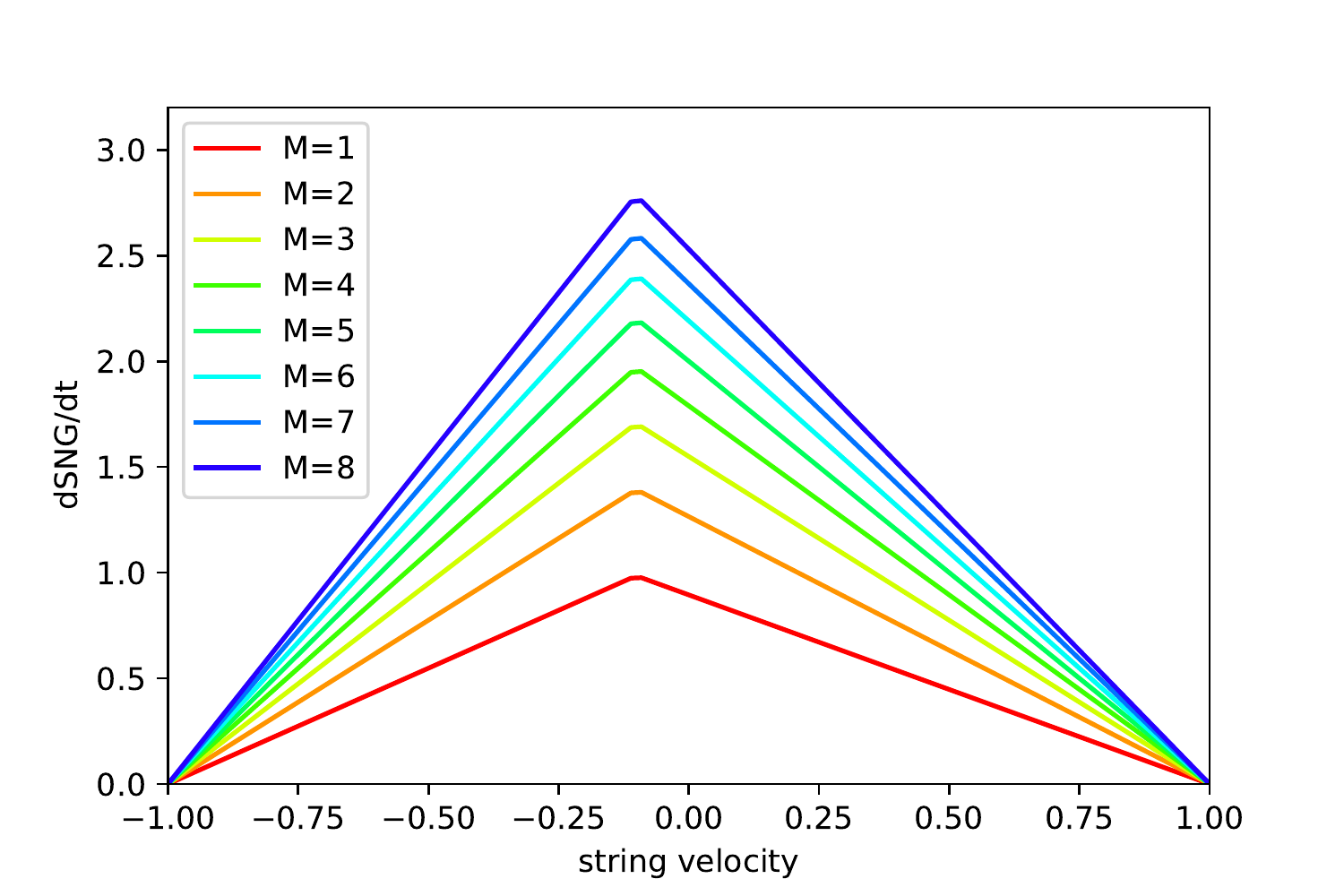}
	\caption{BTZ: Action growth - string velocity $J/M = - 0.2$ fixed}
	\label{fig:BTZActionvelocityJM02}
	\end{minipage}
\end{figure}

\begin{figure}[h]
	\begin{minipage}[t]{0.5\linewidth}
	\includegraphics[width=\linewidth]{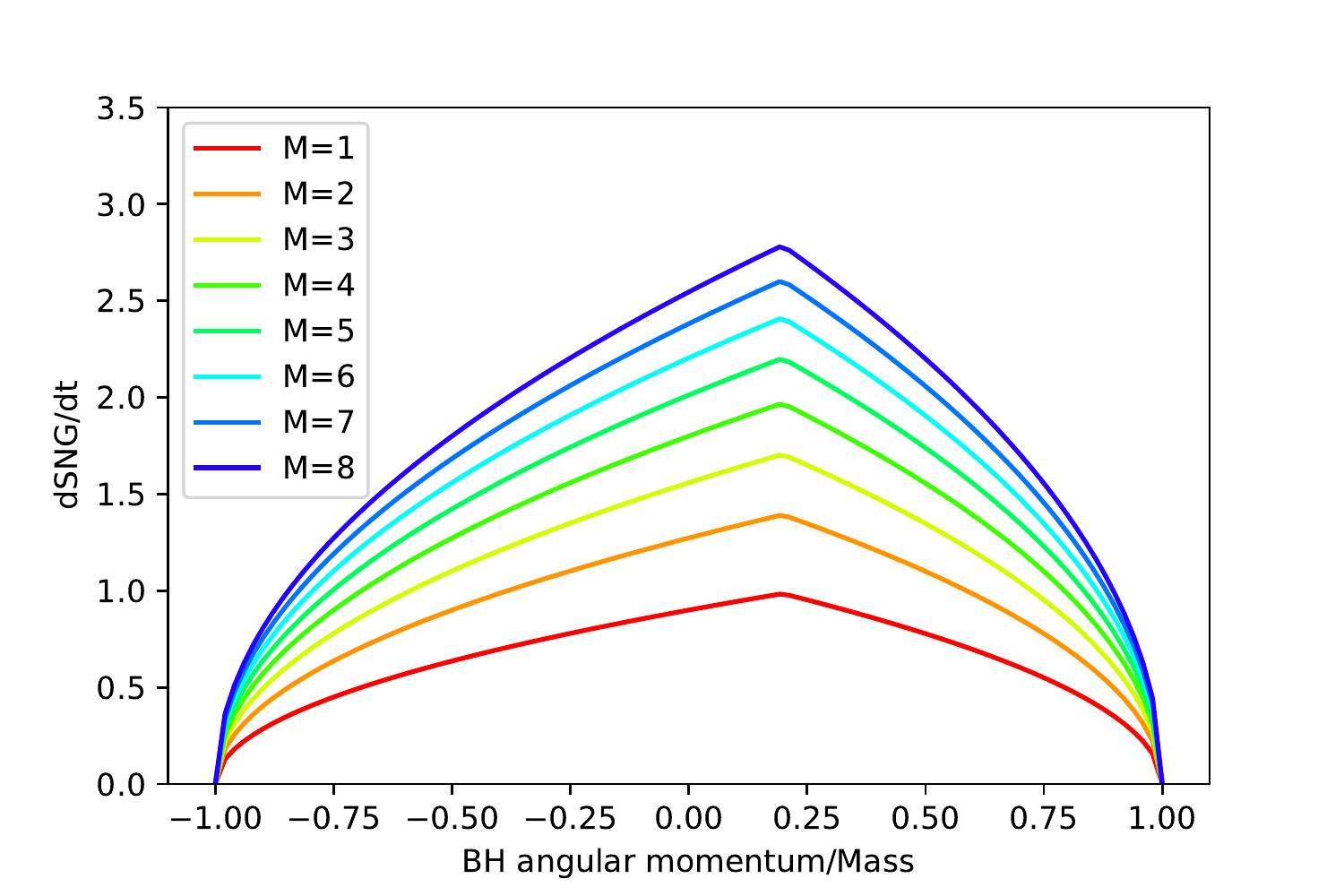}
	\caption{BTZ: Action growth - Black hole angular momentum/Mass}
	\label{fig:BTZActionJM}
	\end{minipage}
\hspace{0.01\linewidth}
	\begin{minipage}[t]{0.5\linewidth}
	\includegraphics[width=\linewidth]{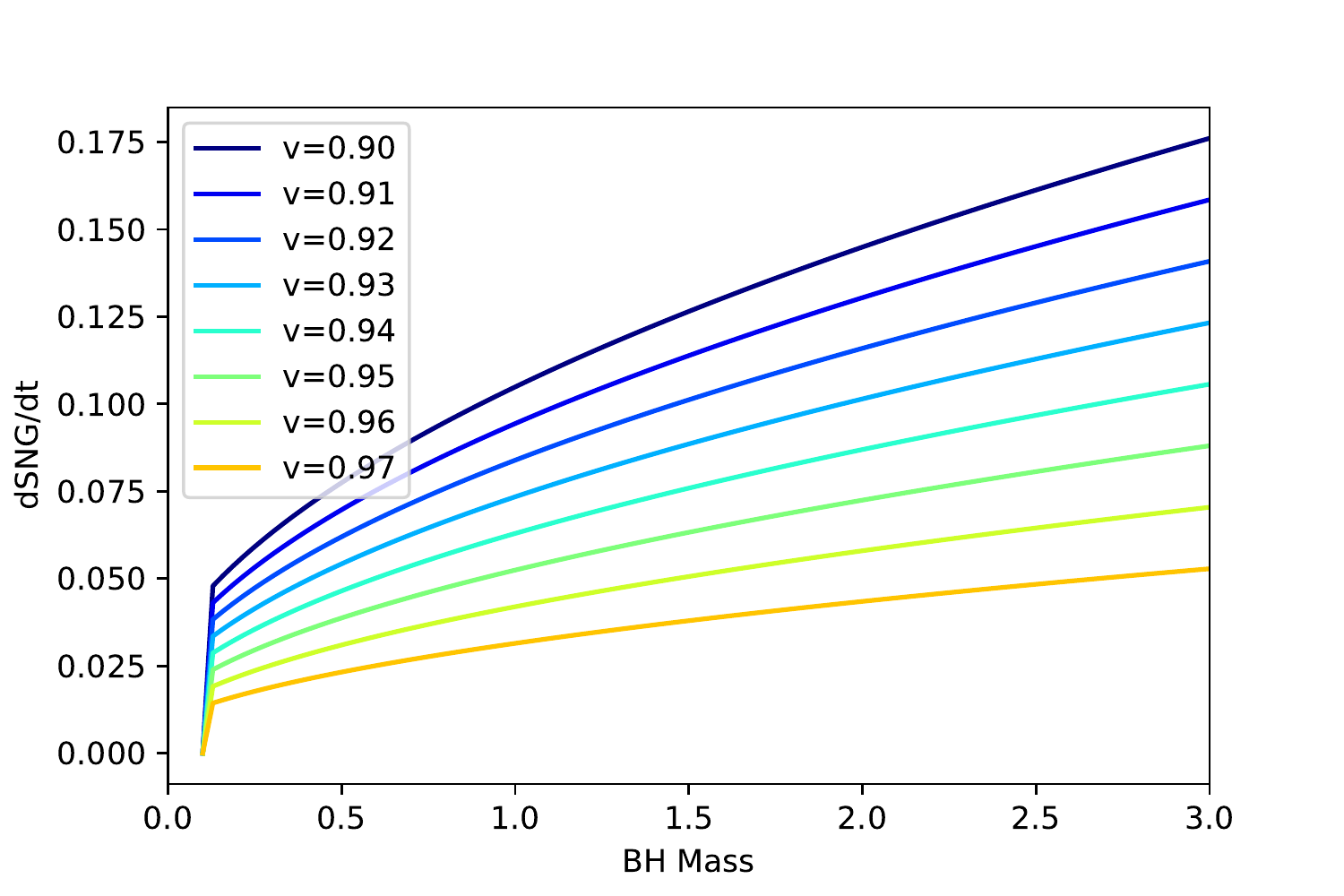}
	\caption{BTZ: Action growth - Black hole mass (small mass region)}
	\label{fig:BTZActionMass2}
	\end{minipage}
\end{figure}

\paragraph{Action}
The development of the Nambu-Goto action is obtained by integrating this Lagrangian over the WDW patch, 
\begin{equation}
\frac{1}{T_\text{s}}\frac{dS_\text{NG}}{dt}
= \sqrt{1-v^2}\int_{r_{-}}^{r_{+}} d\sigma
  \frac{\sigma}{\sqrt{\sigma^2 + \sigma_\text{H}^2 - M}}
= |r_{+} - vr_{-}| - |r_{-} - vr_{+}|.
\end{equation}
In our scaling, $M = r_{+}^2+r_{-}^2,\; J = 2r_{+}r_{-}$, 
$r_\pm = \frac12(\sqrt{M + J}\pm\sqrt{M - J})$.
We express the above action by the parameter $M$ and $J$.
\begin{align}
\frac{1}{T_\text{s}}\frac{dS_\text{NG}}{dt}
= \frac12\big((1-v)\sqrt{M+J} + (1+v)\sqrt{M-J}\big)
- \frac12\big|(1-v)\sqrt{M+J} - (1+v)\sqrt{M-J}\big|.\label{eq:BTZdAdt}
\end{align}
This plot is shown in figures \ref{fig:BTZActionvelocityJM09}, \ref{fig:BTZActionvelocityJM02}, \ref{fig:BTZActionJM} and \ref{fig:BTZActionMass2}.

\begin{figure}[h]
	\begin{minipage}[t]{0.5\linewidth}
	\includegraphics[width=\linewidth]{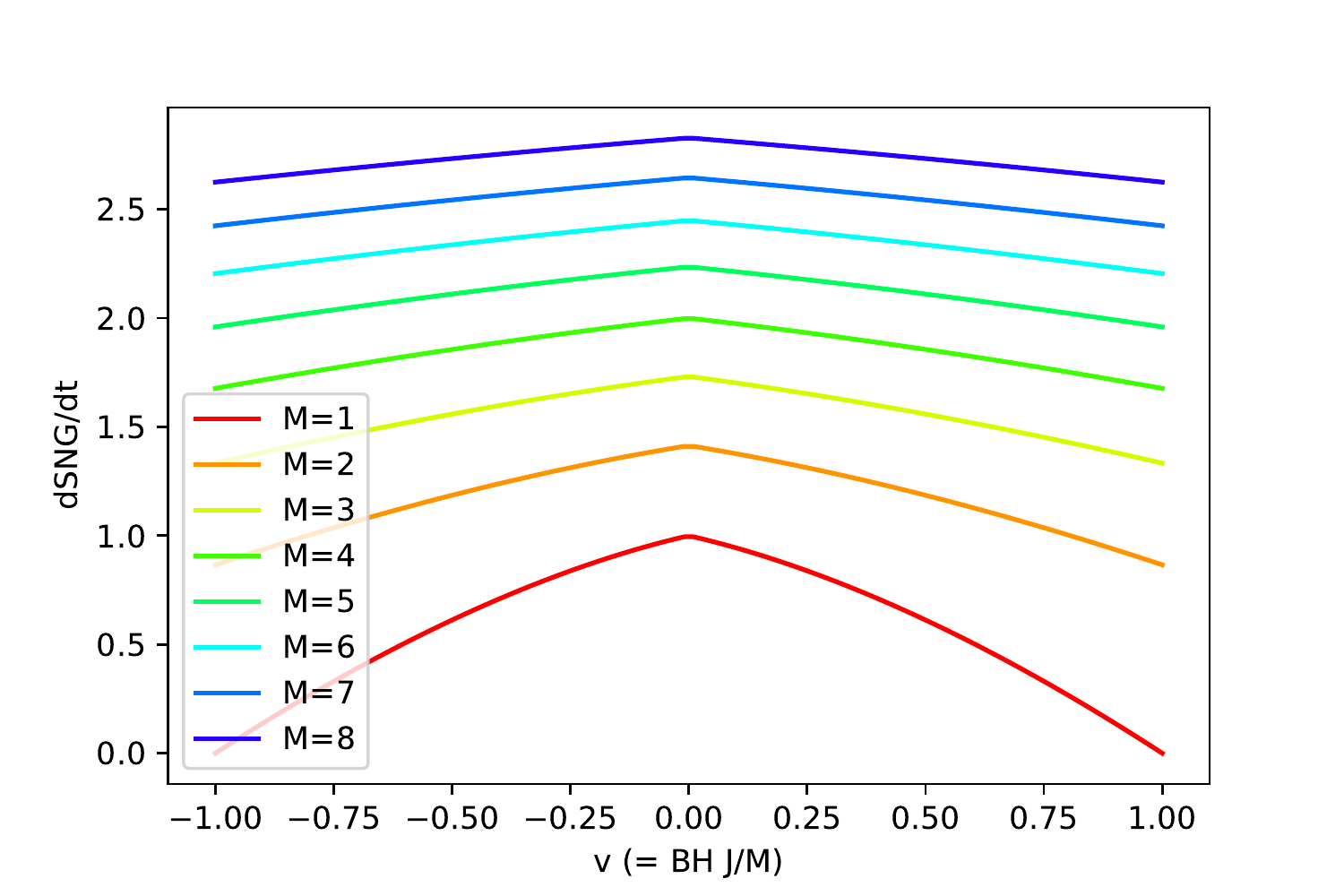}
	\caption{BTZ: Action growth - string velocity (relative velocity$=0$)}
	\label{fig:BTZActionRelv0}
	\end{minipage}
\hspace{0.01\linewidth}
	\begin{minipage}[t]{0.5\linewidth}
	\includegraphics[width=\linewidth]{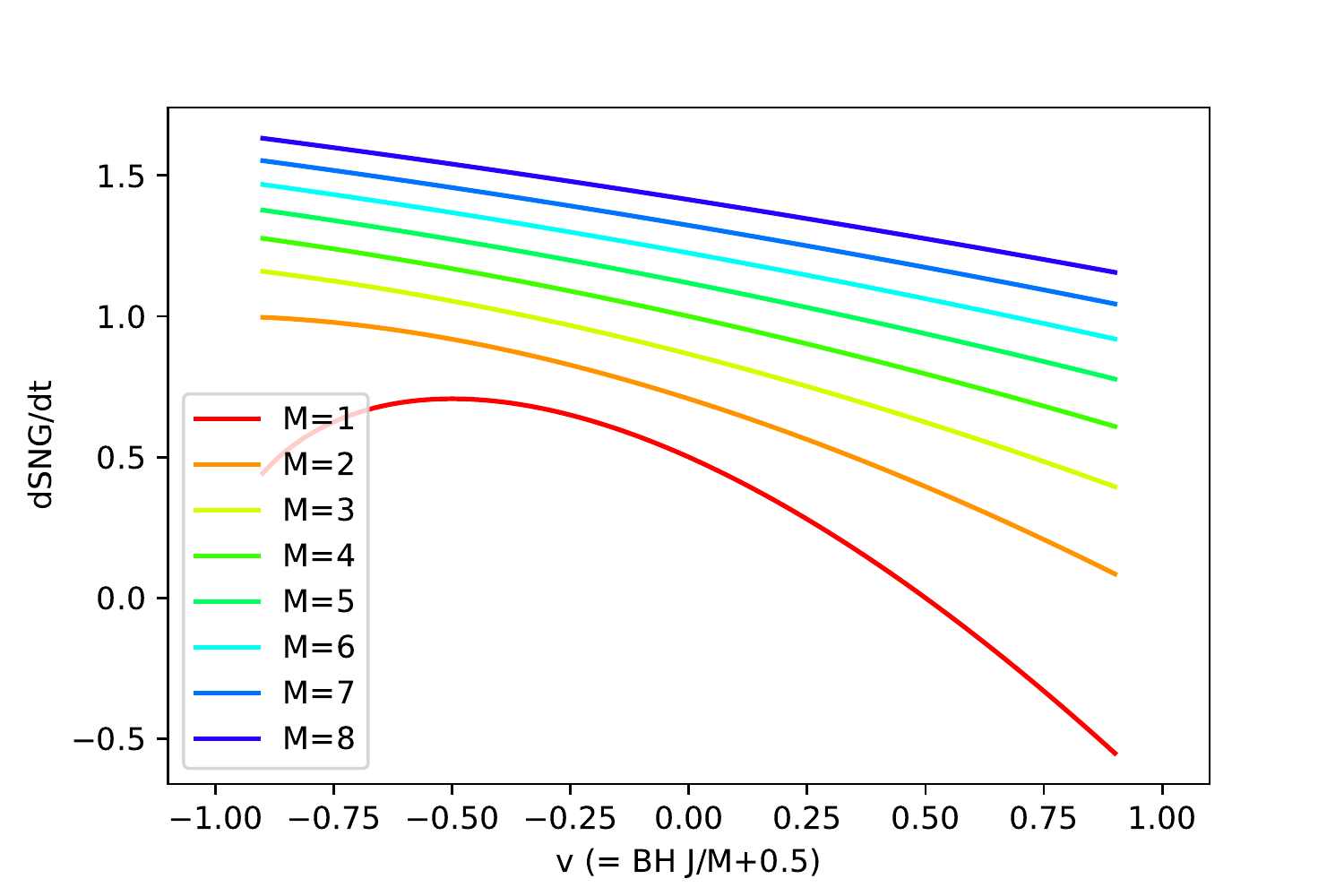}
	\caption{BTZ: Action growth - string velocity (relative velocity$=0.5$)}
	\label{fig:BTZActionRelv05}
	\end{minipage}
\end{figure}

According to the results in the previous work \cite{Nagasaki:2017kqe}, complexity growth is expected to take the maximum when the string is stationary.
So it seems to be meaningful to see the dependence of the relative velocity between the black hole and the string.
That dependence of the relative velocity is shown in figures \ref{fig:BTZActionRelv0} and \ref{fig:BTZActionRelv05}.
As expected, this tendency is seen in these plots.
That is, the effect to complexity takes maximum when the relative velocity is zero for rotating black holes.
The effect is larger for larger black holes as AdS$_4$ case \cite{Nagasaki:2017kqe}.

Figure \ref{fig:BTZActionvelocityJM09} and figure \ref{fig:BTZActionvelocityJM02} show
the velocity dependence for different angular momentums $J/M$.
In the left figure \ref{fig:BTZActionvelocityJM09} the peak position is shifted because of the black hole rotation.
In the right figure \ref{fig:BTZActionvelocityJM02}, since the black hole rotates in the opposite direction to the string, the peak is shifted in the opposite side.
Not only the peak position is shifted, we can also see that the peak value becomes smaller for a large shift.
Note that in this case the effect of the probe string is exactly zero when the string motion reaches to the light speed.

Figure \ref{fig:BTZActionMass2} also shows that the effect of the string is smaller for fast moving strings.
And this is a monotonically increasing function of the black hole mass.
This can be thought that this is because complexity defines how complex of the physical system.
Then a larger system may have larger information.


\section{Kerr-AdS black holes}\label{sec:KABH}
In this section we consider the Kerr-AdS black holes which are black hole solutions with angular momentum.
General higher dimensional solution is known and that holographic correspondence is studied in \cite{Awad:2000aj, Lu:2008jk}. 

\subsection{Four dimensional Kerr-AdS black holes}
We consider here AdS black holes with angular momentum.
The Einstein-Hilbert action of this black holes is studied in \cite{Cai:2016xho}.
We study an effect of the probe string here.
In the Boyer-Lindquist coordinates the four-dimensional Kerr-AdS black hole is given by
\begin{align}\label{eq:dsKA4}
ds_\text{KA4}^2 
&= -\frac{\Delta_r}{\rho^2}\Big(dt-\frac{a\sin^2\theta}{\Xi}d\phi\Big)^2
 + \frac{\rho^2}{\Delta_r}dr^2 + \frac{\rho^2}{\Delta_\theta}d\theta^2
 + \frac{\Delta_\theta\sin^2\theta}{\rho^2}\Big(adt-\frac{r^2+a^2}{\Xi}d\phi\Big)^2\\
&= -\Big(\frac{\Delta_r}{\rho^2}-\frac{\Delta_\theta\sin^2\theta}{\rho^2}a^2\Big)dt^2
 + \frac{\rho^2}{\Delta_r}dr^2 + \frac{\rho^2}{\Delta_\theta}d\theta^2\nonumber\\
&\qquad
 + 2\frac{a\Delta_r\sin^2\theta-a(r^2+a^2)\Delta_\theta\sin^2\theta}{\rho^2\Xi}dtd\phi
 + \frac{(r^2+a^2)^2\Delta_\theta\sin^2\theta-a^2\Delta_r\sin^4\theta}{\rho^2\Xi^2}d\phi^2,\;
\end{align}
where
\begin{align*}\label{eq:dsKAprs}
\Delta_r &= (r^2+a^2)(1+r^2/\ell_\text{AdS}^2)-2mr,\;
\Xi = 1-a^2/\ell_\text{AdS}^2,\\
\Delta_\theta &= 1-a^2\cos^2\theta/\ell_\text{AdS}^2,\;
\rho^2 = r^2 + a^2\cos^2\theta.
\end{align*}
The physical mass and the angular momentum are
$M = m/(G_\text{N}\Xi)$ and $J = ma/(G_\text{N}\Xi)$.
The above metric is related with the AdS coordinates $\Phi$ (see also (4.11) of \cite{Hawking:1998kw}) as follows.
Put the AdS boundary coordinates as $t$ and $\Omega$ and they are related by $\Phi = \phi + \Omega t$.
In this coordinate, the first and the last terms of eq.\eqref{eq:dsKA4} are, at $\theta=\pi/2$,
\begin{align*}
-r^2\ell_\text{AdS}^{-2}
  \Big[dt - \frac{a}{\Xi}d\phi\Big]^2
&= -r^2\ell_\text{AdS}^{-2}
  \Big[\Big(1-\frac{a\Omega}{\Xi}\Big)dt - \frac{a}{\Xi}d\Phi\Big]^2,\\
\frac{1}{r^2}
  \Big[adt - \frac{r^2}{\Xi}d\phi\Big]^2
&= \frac{1}{r^2}
  \Big[\Big(a-\frac{r^2\Omega}{\Xi}\Big)dt - \frac{r^2}{\Xi} d\Phi\Big]^2.
\end{align*}
In order for these terms to give the form of AdS metric at $r\rightarrow\infty$, the cross terms from these terms should cancel. Then, the parameter $\Omega$ is determined 
\begin{equation}
\phi = \Phi - a\ell_\text{AdS}^{-2}t.\label{eq:Phishift}
\end{equation}

We consider a string moving in this spacetime.
In the following we use the rescaled coordinates so that $\ell_\text{AdS} = 1$.
We assume that the string moves on the great circle of the subspace $S^2$: $\theta=\pi/2$.
We parametrize the string worldsheet as
\footnote{
In this section we use the capital letters $\Phi$ and $V$ for boundary coordinate and the string velocity.
These are shifted because of the rotation of the black hole.
}
\begin{equation}
t = \tau,\; r = \sigma,\; \Phi = V\tau + \xi(\sigma).
\end{equation}
Taking into account the relation \eqref{eq:Phishift}, the above is 
\begin{equation}
t = \tau,\; r = \sigma,\; 
\phi = (V-a)\tau + \xi(\sigma)
:= v\tau + \xi(\sigma).
\end{equation}
We defined the shifted velocity as $v:=V-a$ which is used in the following calculation while the original $V$ is the string velocity.
Let us define a function
\begin{equation}
\Delta(\sigma) := \Delta_r(r=\sigma)
= (\sigma^2+a^2)(1+\sigma^2)-2m\sigma.
\end{equation}
The induced metric is 
\begin{align}
ds_\text{KA4ind}^2 
&= -\Big(\frac{\Delta}{\sigma^2}-\frac{a^2}{\sigma^2}
  - 2\frac{av}{\sigma^2}\frac{\Delta - (\sigma^2+a^2)}{\Xi}
  - v^2\frac{(\sigma^2+a^2)^2-a^2\Delta}{\sigma^2\Xi^2}\Big)d\tau^2\nonumber\\
&\qquad
 + \Big(\frac{\sigma^2}{\Delta} + \frac{(\sigma^2+a^2)^2-a^2\Delta}{\sigma^2\Xi^2}\xi'^2\Big)d\sigma^2\nonumber\\
&\qquad
 + \frac{2\xi'}{\sigma^2\Xi}\Big(a(\Delta-(\sigma^2+a^2)) + v\frac{(\sigma^2+a^2)^2-a^2\Delta}{\Xi}\Big)d\tau d\sigma.
\end{align}
We define functions for simplicity:
\begin{equation}
F := \Delta - (\sigma^2 + a^2),\;
G := (\sigma^2 + a^2)^2 - a^2\Delta.
\end{equation}
Its determinant is 
\begin{align}
-\det[g_\text{KA4ind}]
&= \Big(\frac{\Delta}{\sigma^2} - \frac{a^2}{\sigma^2}
  - 2\frac{av}{\sigma^2}\frac{F}{\Xi}
  - v^2\frac{G}{\sigma^2\Xi^2}\Big)
\Big(\frac{\sigma^2}{\Delta} + \frac{G}{\sigma^2\Xi^2}\xi'^2\Big)
 + \frac{\xi'^2}{\sigma^4\Xi^2}\Big(aF + v\frac{G}{\Xi}\Big)^2.
\end{align}
Further we define
\begin{equation}
H(\sigma) := \frac{\Delta}{\sigma^2} - \frac{a^2}{\sigma^2}
  - 2\frac{av}{\sigma^2}\frac{F}{\Xi}
  - v^2\frac{G}{\sigma^2\Xi^2},\;
I(\sigma) := \Big(aF + v\frac{G}{\Xi}\Big)^2.
\end{equation}
\paragraph{EOM and Lagrangian}
The Lagrangian and the equation of motion is
\begin{equation}
\frac{\mathcal L_\text{KA4}}{T_\text{s}}
= \Big[H\Big(\frac{\sigma^2}{\Delta} + \frac{G}{\sigma^2\Xi^2}\xi'^2\Big)
 + \frac{\xi'^2}{\sigma^4\Xi^2}I\Big]^{1/2},\;\;
\frac{1}{T_\text{s}}\frac{\partial\mathcal L_\text{KA4}}{\partial\xi'(\sigma)}
= \frac{\xi'/(\sigma^4\Xi^2)}{\mathcal L_\text{KA4}/T_\text{s}}
  [\sigma^2HG + I]
=: c_\xi.
\end{equation}
Solving it for $\xi'(\sigma)$, we obtain
\begin{equation}
\xi' = c_\xi\sigma^4\Xi^2
  \sqrt{\frac{\sigma^2H/\Delta}{(\sigma^2HG+I)(\sigma^2HG+I-c_\xi^2\sigma^4\Xi^2)}}.
\end{equation}
We impose the reality condition as before.
We need to find the zero of
\begin{align}
\sigma^2H 
&= \Delta-a^2 - 2av\frac{\Delta-(\sigma^2+a^2)}{\Xi}
  - v^2\frac{(\sigma^2+a^2)^2-a^2\Delta}{\Xi^2}\nonumber\\
&= \Delta(\sigma)\Big(1-\frac{av}{\Xi}\Big)^2
  - \Big(a - \frac{v(\sigma^2+a^2)}{\Xi}\Big)^2.
\end{align}
We denote this solution as $\sigma = \sigma_\text{H}$.
Actually $\sigma^2H(\sigma) = 0$ has a unique solution at positive region. 
The denominator must be zero coincidentally.
This condition determines the integration constant $c_\xi$:
\begin{equation}
0 = I(\sigma_\text{H})^2 - c_\xi^2\Xi^2\sigma_\text{H}^4I(\sigma_\text{H}),\;\;
\qquad\therefore 
c_\xi^2\Xi^2 = I(\sigma_\text{H})/\sigma_\text{H}^4.
\end{equation}
Substituting this constant into the above, we obtain
\begin{equation}
\xi'(\sigma) = c_\xi\sigma^4\Xi^2
  \sqrt{\frac{\sigma^2H(\sigma)/\Delta(\sigma)}{(\sigma^2H(\sigma)G(\sigma)+I(\sigma))(\sigma^2H(\sigma)G(\sigma) + I(\sigma) - I(\sigma_\text{H})\sigma^4/\sigma_\text{H}^4)}},
\end{equation}
and the Lagrangian
\begin{equation}
\frac{\mathcal L_\text{KA4}}{T_\text{s}}
= \frac{[\sigma^2HG + I]}{c_\xi\sigma^4\Xi^2}\xi'(\sigma)
= \sqrt{\frac{(\sigma^2H(\sigma)G(\sigma)+I(\sigma))\sigma^2H(\sigma)/\Delta(\sigma)}{\sigma^2H(\sigma)G(\sigma) + I(\sigma) - I(\sigma_\text{H})\sigma^4/\sigma_\text{H}^4}}.\label{eq:LagKA4}
\end{equation}

\begin{figure}[h]
	\begin{minipage}[t]{0.5\linewidth}
	\includegraphics[width=\linewidth]{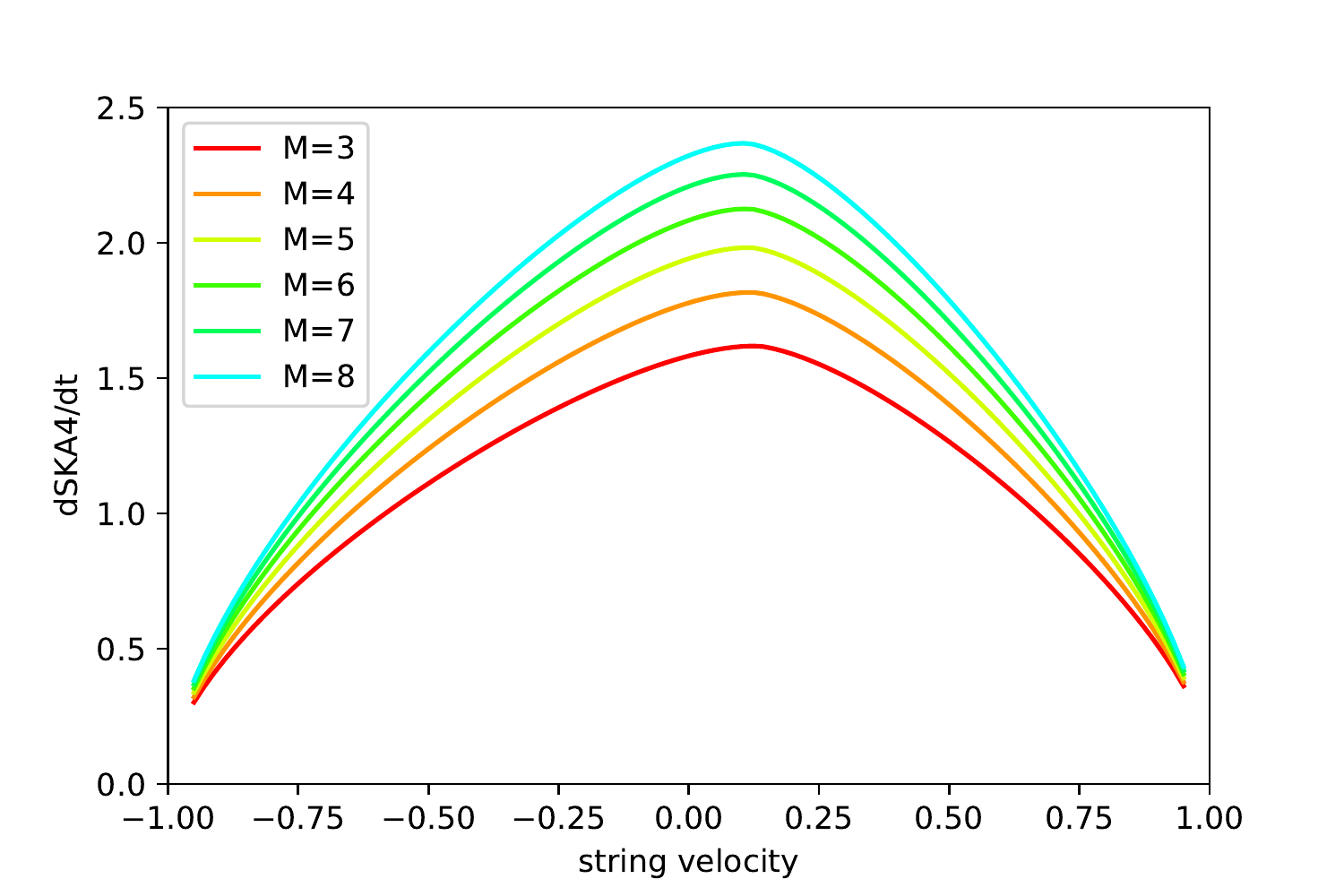}
	\caption{Kerr-AdS$_{3+1}$: Action growth - string velocity ($a = 0.1$)}
	\label{fig:KA4Actvelocitya01}
	\end{minipage}
\hspace{0.01\linewidth}
	\begin{minipage}[t]{0.5\linewidth}
	\includegraphics[width=\linewidth]{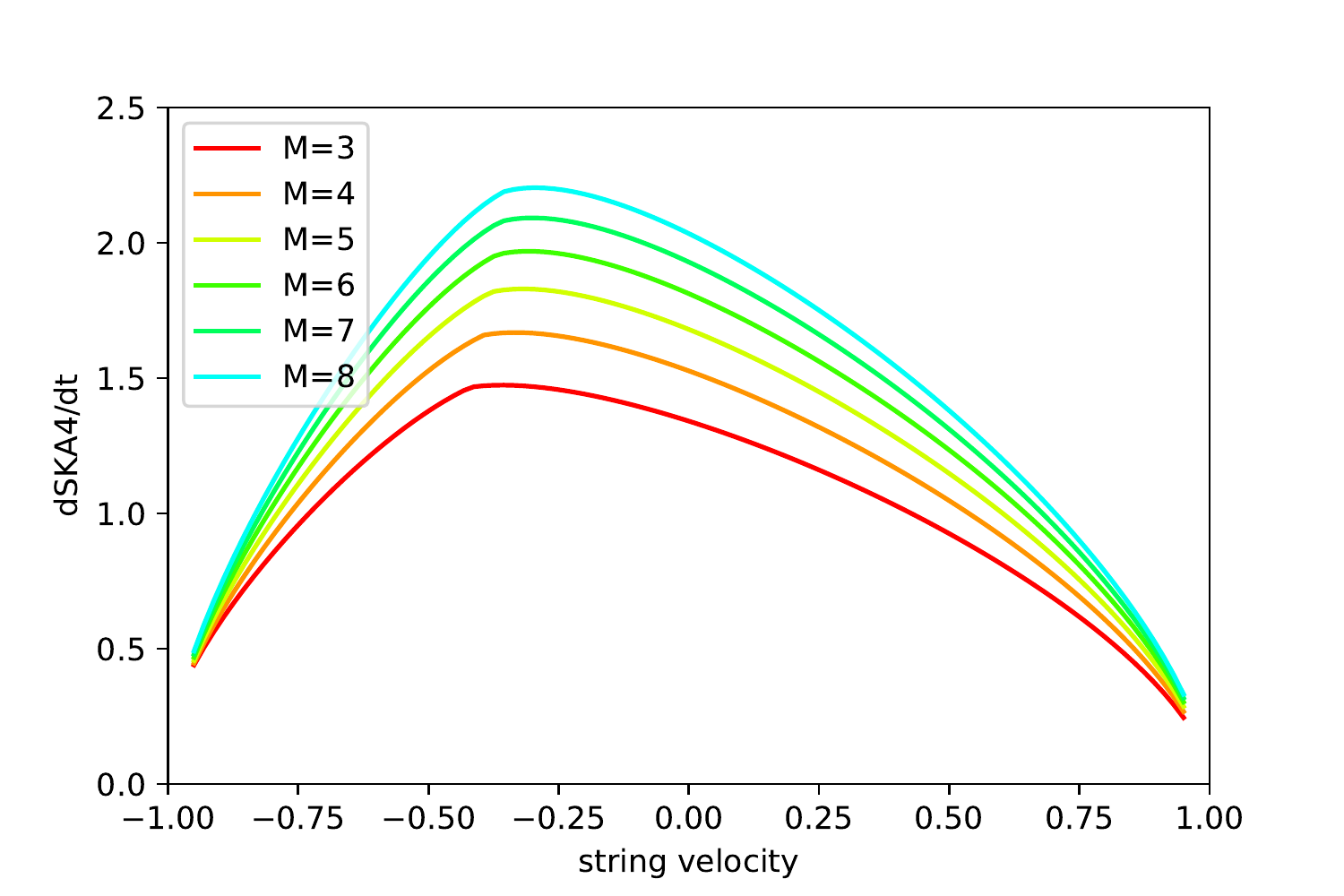}
	\caption{Kerr-AdS$_{3+1}$: Action growth - string velocity ($a = -0.2$)}
	\label{fig:KA4Actvelocityan02}
	\end{minipage}
\end{figure}

\begin{figure}[h]
	\begin{minipage}[t]{0.5\linewidth}
	\includegraphics[width=\linewidth]{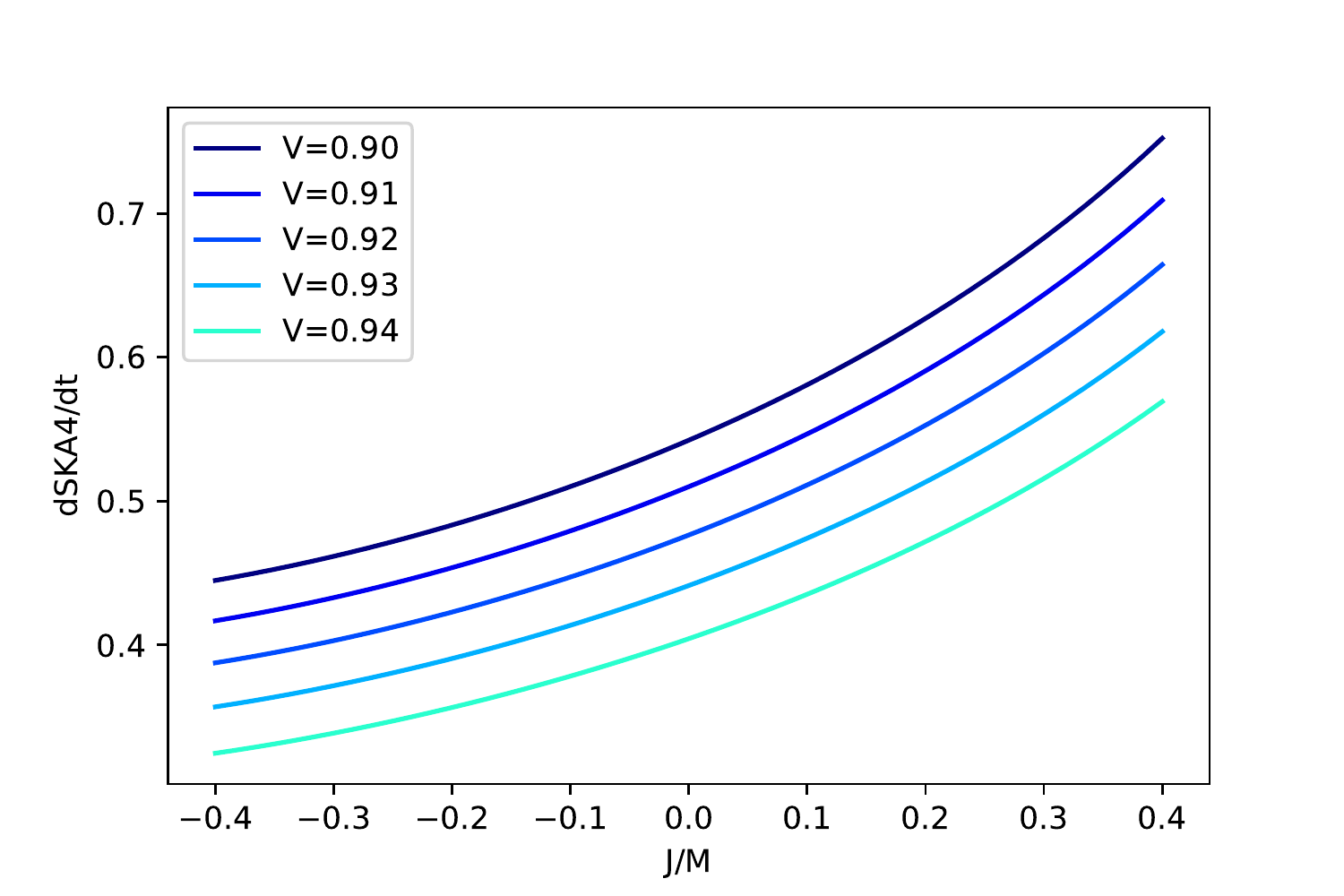}
	\caption{Kerr-AdS$_{3+1}$: Action growth - Black hole $J/M$ ($m=5$)}
	\label{fig:KA4ActionJM}
	\end{minipage}
\hspace{0.01\linewidth}
	\begin{minipage}[t]{0.5\linewidth}
	\includegraphics[width=\linewidth]{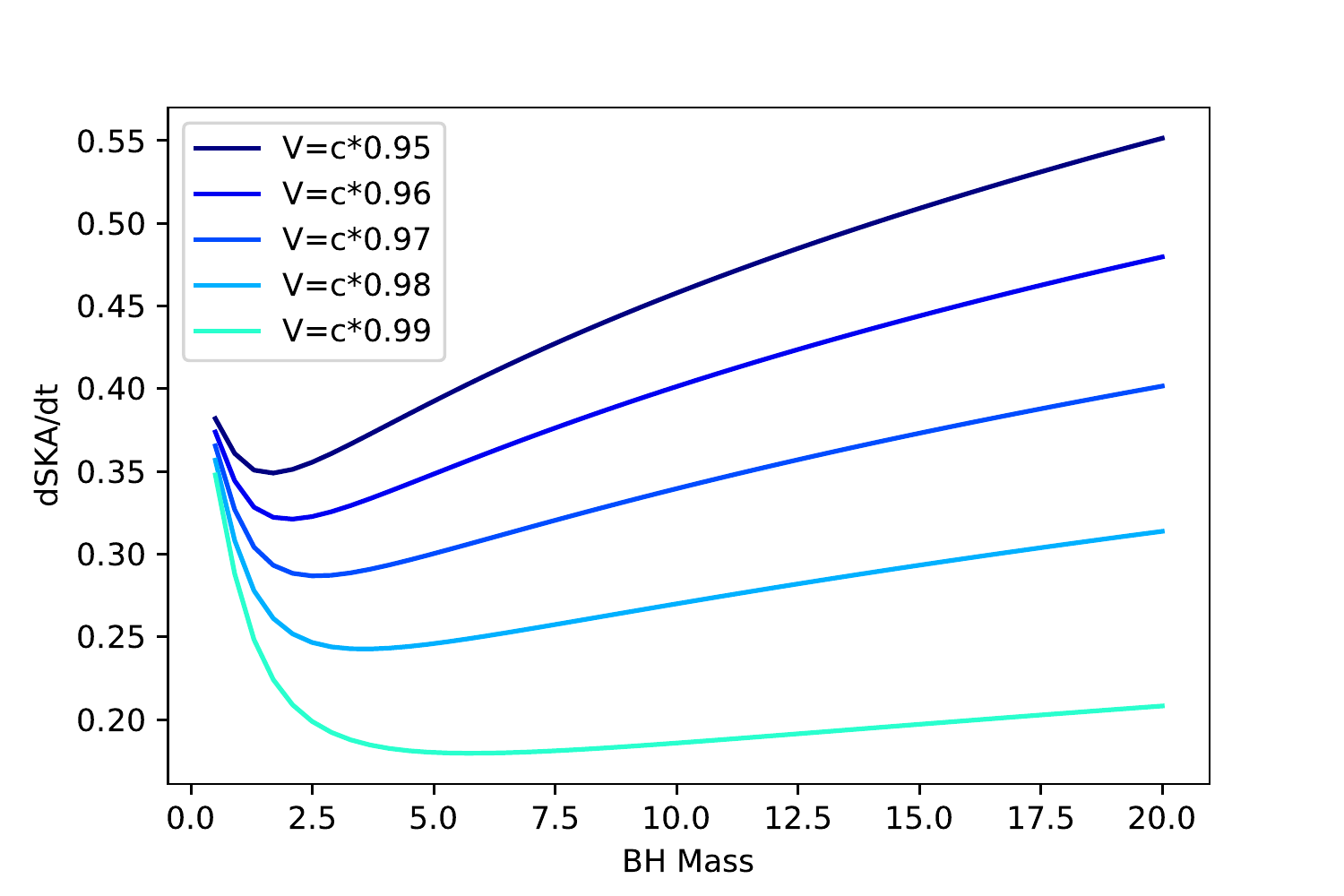}
	\caption{Kerr-AdS$_{3+1}$: Action growth - Black hole mass ($a=0.1$)}
	\label{fig:KA4ActionMass}
	\end{minipage}
\end{figure}

\paragraph{Action}
The outer and the inner horizon are the smaller and the larger solutions of $\Delta_r(r) = 0$, respectively.
The integral in the WDW patch is 
\begin{align}
\frac{1}{T_\text{s}}\int_{r_{-}}^{r_{+}} d\sigma\mathcal L_\text{KA4}
= \int_{r_{-}}^{r_{+}} d\sigma
 \sqrt{\frac{\sigma^2H(\sigma)(\sigma^2H(\sigma)G(\sigma)+I(\sigma))}{\Delta(\sigma)(\sigma^2H(\sigma)G(\sigma) + I(\sigma) - I(\sigma_\text{H})\sigma^4/\sigma_\text{H}^4)}}.
\end{align}
The numerical calculation gives the results shown in figures \ref{fig:KA4Actvelocitya01}, \ref{fig:KA4Actvelocityan02}, \ref{fig:KA4ActionJM} and \ref{fig:KA4ActionMass}.

The first two figures, figure \ref{fig:KA4Actvelocitya01} and figure \ref{fig:KA4Actvelocityan02}, show the string velocity dependence in different masses.
The left one is the result for $a=0.1$.
The peak position is shifted to the right side.
The right figure is, on the other hand, the result for a black hole with angular momentum of  the opposite direction $a= - 0.2$.
The peak position is shifted to the other side.
This behavior is consistent with the property that complexity is larger as the probe string motion is slower.
That is, the effect takes the peak value when the relative velocity between the string and the black hole is zero.
Not only that we can also the peak value decreases as the shift becomes larger similar to the BTZ black hole case.

Figure \ref{fig:KA4ActionJM} is the dependence of the angular momentum per black hole mass.
This figure also shows that the effect to complexity is smaller as the string moves faster.
And It also shows the tendency that complexity is larger when the relative velocity is smaller because in this plot the string velocity is positive and the effect of complexity is larger in the positive region of the graph.

Figure \ref{fig:KA4ActionMass} shows the mass dependence.
As usual this is a increasing function of black hole mass.
The faster string gives the small effect to complexity.
But there is an unusual behavior in the small mass region where the extremum value appears.


\subsection{Five dimensional Kerr-AdS black holes}
The (4+1)-dimensional Kerr-AdS black hole is described by (see references \cite{Hawking:1998kw, Hawking:1999dp, Lu:2008jk, Tsai:2011gv})
\begin{align}
ds_\text{KA5}^2
&= -\frac{\Delta_r}{\rho^2}
  \Big(dt-\frac{a\sin^2\theta}{\Xi_a}d\phi_1 - \frac{b\cos^2\theta}{\Xi_b}d\phi_2\Big)^2\nonumber\\
& + \frac{\Delta_\theta\sin^2\theta}{\rho^2}
  \Big(adt-\frac{r^2+a^2}{\Xi_a}d\phi_1\Big)^2
 + \frac{\Delta_\theta\cos^2\theta}{\rho^2}
  \Big(bdt-\frac{r^2+b^2}{\Xi_b}d\phi_2\Big)^2\nonumber\\
& + \frac{\rho^2}{\Delta_r}dr^2 + \frac{\rho^2}{\Delta_\theta}d\theta^2
 + \frac{1+r^2}{r^2\rho^2}
   \Big(abdt 
  - \frac{b(r^2+a^2)\sin^2\theta}{\Xi_a}d\phi_1
  - \frac{a(r^2+b^2)\cos^2\theta}{\Xi_b}d\phi_2\Big)^2,
\end{align}
where
\begin{align*}
&
\rho^2(r) = r^2 + a^2\cos^2\theta + b^2\sin^2\theta,\\
&
\Delta_r(r) = \frac1{r^2}(r^2+a^2)(r^2+b^2)(r^2+1) - 2m,\qquad
\Delta_\theta(\theta) = 1 - a^2\cos^2\theta - b^2\sin^2\theta,\\
&
\Xi_a = 1-a^2,\qquad
\Xi_b = 1-b^2.
\end{align*}
The parameters here is related to the physical mass and the angular momentum as (\cite{Hawking:1999dp})
\begin{equation}\label{eq:KA5relMJ}
M = \frac{3\pi m}{4\Xi_1\Xi_2},\;
J_i = \frac{\pi a_im}{2\Xi_i(1+r_{+}^2)}.
\end{equation}
We study the $a\neq 0$ case and the $b\neq 0$ case separately.
These correspond to black holes rotating around different axises with coordinates $\phi_1$ and $\phi_2$.  
As before we assume that the string moves in the great circle: $\theta = \pi/2$.

\subsubsection{$a\neq 0$ case}
First we consider the case where only $a$ is nonzero.
In this case the string rotates around the same axis to the black hole.
\begin{subequations}
\begin{align}
ds_\text{KA5a}^2
&= -\frac{\Delta_{ra}}{r^2}
  \Big(dt-\frac{a}{\Xi_a}d\phi_1\Big)^2
 + \frac{r^2}{\Delta_{ra}}dr^2 
 + \frac{1}{r^2}
  \Big(adt-\frac{r^2+a^2}{\Xi_a}d\phi_1\Big)^2,\\
\Delta_{ra} &= (r^2+a^2)(r^2+1) - 2m.
\end{align}
\end{subequations}
This looks the same form to the four-dimensional Kerr-AdS case except that the function $\Delta_r(r)$ is replaced with $\Delta_{ra}(r)$ (the second term does not depend on $r$).
We needs the same shift \eqref{eq:Phishift} to relate the velocity parameter $v$ to the string velocity $V$: $v = V-a$
The parametrization of the string worldsheet is 
\begin{equation}
t = \tau,\;
r = \sigma,\;
\Phi = V\tau + \xi(\sigma).
\end{equation}

\paragraph{EOM and its solution}
The calculation of the induced metric and the NG action are performed in the same way as the Kerr-AdS$_{3+1}$ case.
So the Lagrangian is the same form to \eqref{eq:LagKA4} except that $\Delta(r)$ is replaced with $\Delta_a$: 
\begin{equation}
\frac{\mathcal L_\text{KA5a}}{T_\text{s}}
= \sqrt{\frac{(\sigma^2H(\sigma)G(\sigma)+I(\sigma))\sigma^2H(\sigma)/\Delta_a(\sigma)}{\sigma^2H(\sigma)G(\sigma) + I(\sigma) - I(\sigma_\text{H})\sigma^4/\sigma_\text{H}^4}},
\end{equation}
where
\begin{subequations}
\begin{align}
\Delta_a(\sigma) &:= (\sigma^2+a^2)(\sigma^2+1) - 2m,\\
F(\sigma) &:= \Delta_a(\sigma) - (\sigma^2+a^2),\qquad
&
G(\sigma) &:= (\sigma^2+a^2)^2 - a^2\Delta_a(\sigma),\\
H(\sigma) &:= \frac{\Delta_a(\sigma)}{\sigma} - \frac{a^2}{\sigma} - 2\frac{av}{\Xi_a} - v^2\frac{G(\sigma)}{\sigma^2\Xi^2},\qquad
&
I(\sigma) &:= \Big(aF(\sigma) + v\frac{G}{\Xi}\Big)^2.
\end{align}
\end{subequations}
The inner and the outer horizons are determined by 
\begin{equation}
\Delta_{ra}(r_\pm) = 0.
\end{equation}

\paragraph{Action}
The action integrated over the WDW patch is 
\begin{equation}
\frac{dS_\text{NG}}{dt}
= \int_{r_{-}}^{r_{+}}d\sigma
	\sqrt{\frac{(\sigma^2H(\sigma)G(\sigma)+I(\sigma))\sigma^2H(\sigma)/\Delta_a(\sigma)}{\sigma^2H(\sigma)G(\sigma) + I(\sigma) - I(\sigma_\text{H})\sigma^4/\sigma_\text{H}^4}}.
\end{equation}
This integration is performed by the numerical calculation.
The result shows the string velocity dependence, the black hole angular momentum dependence and the mass dependence.

\begin{figure}[h]
	\begin{minipage}[t]{0.5\linewidth}
	\includegraphics[width=\linewidth]{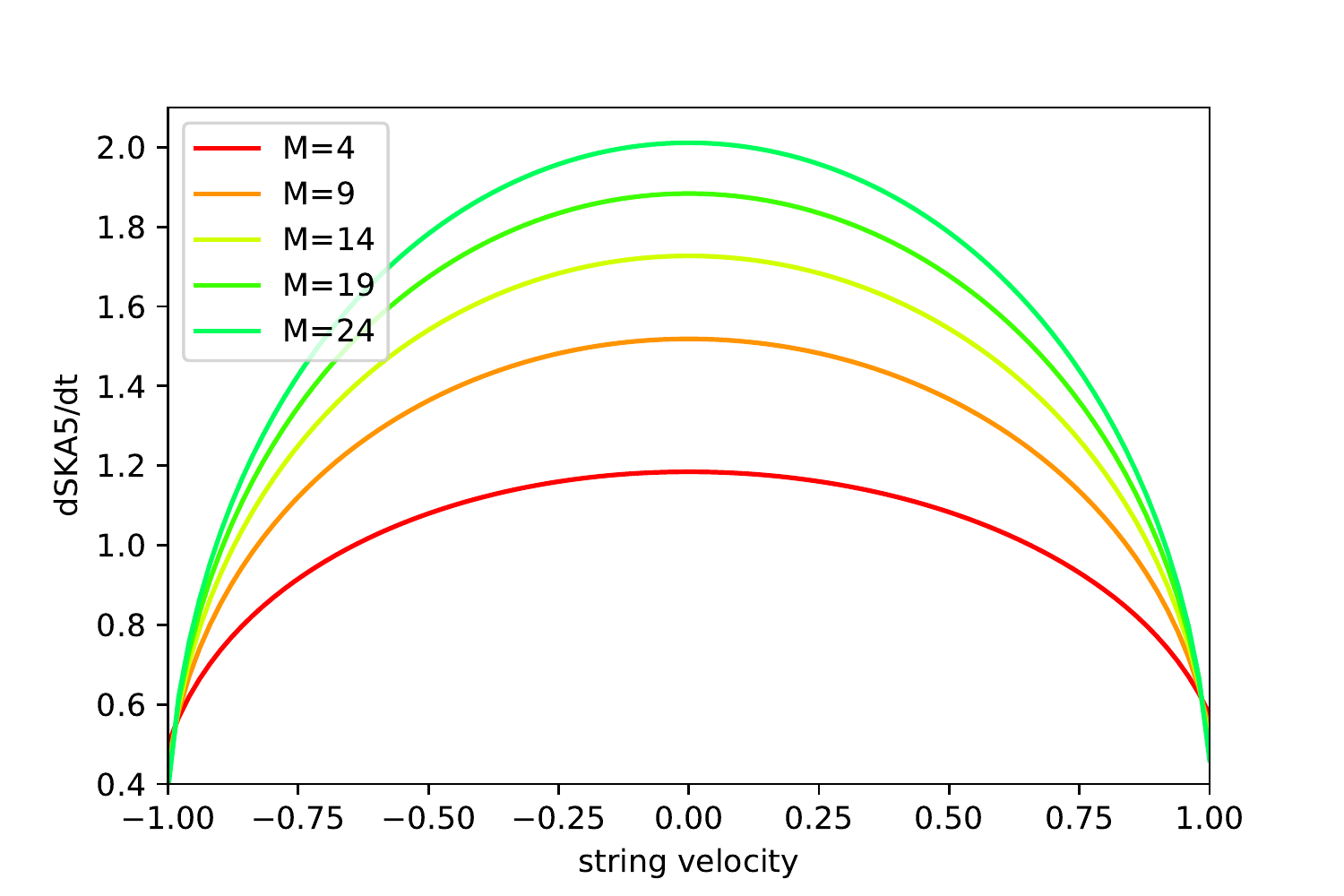}
	\caption{Kerr-AdS$_{4+1}$: Action growth - velocity ($a=0.1$)}
	\label{fig:KA5aSNGvelocitya01}
	\end{minipage}
\hspace{0.01\linewidth}
	\begin{minipage}[t]{0.5\linewidth}
	\includegraphics[width=\linewidth]{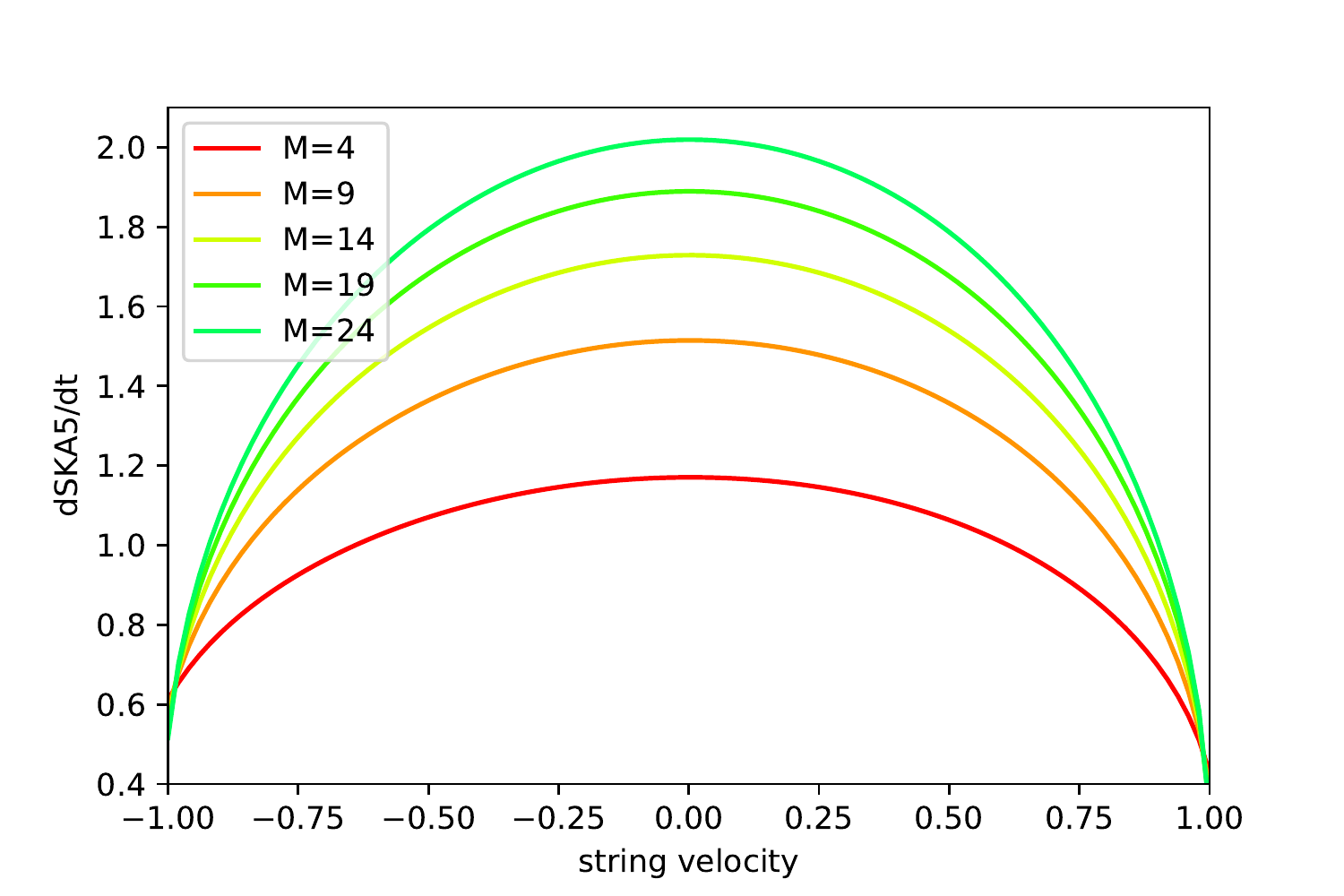}
	\caption{Kerr-AdS$_{4+1}$: Action growth - $m$ ($a= -0.2$)}
	\label{fig:KA5aSNGvelocityan025}
	\end{minipage}
\end{figure}

The velocity dependence is shown in figure \ref{fig:KA5aSNGvelocitya01} and figure \ref{fig:KA5aSNGvelocityan025}.
We can see as usual the effect to complexity is larger for larger mass and the effect is maximum when the string is stationary. 
But the difference for different values of parameter $a$ disappears.

\begin{figure}[h]
	\begin{minipage}[t]{0.5\linewidth}
	\includegraphics[width=\linewidth]{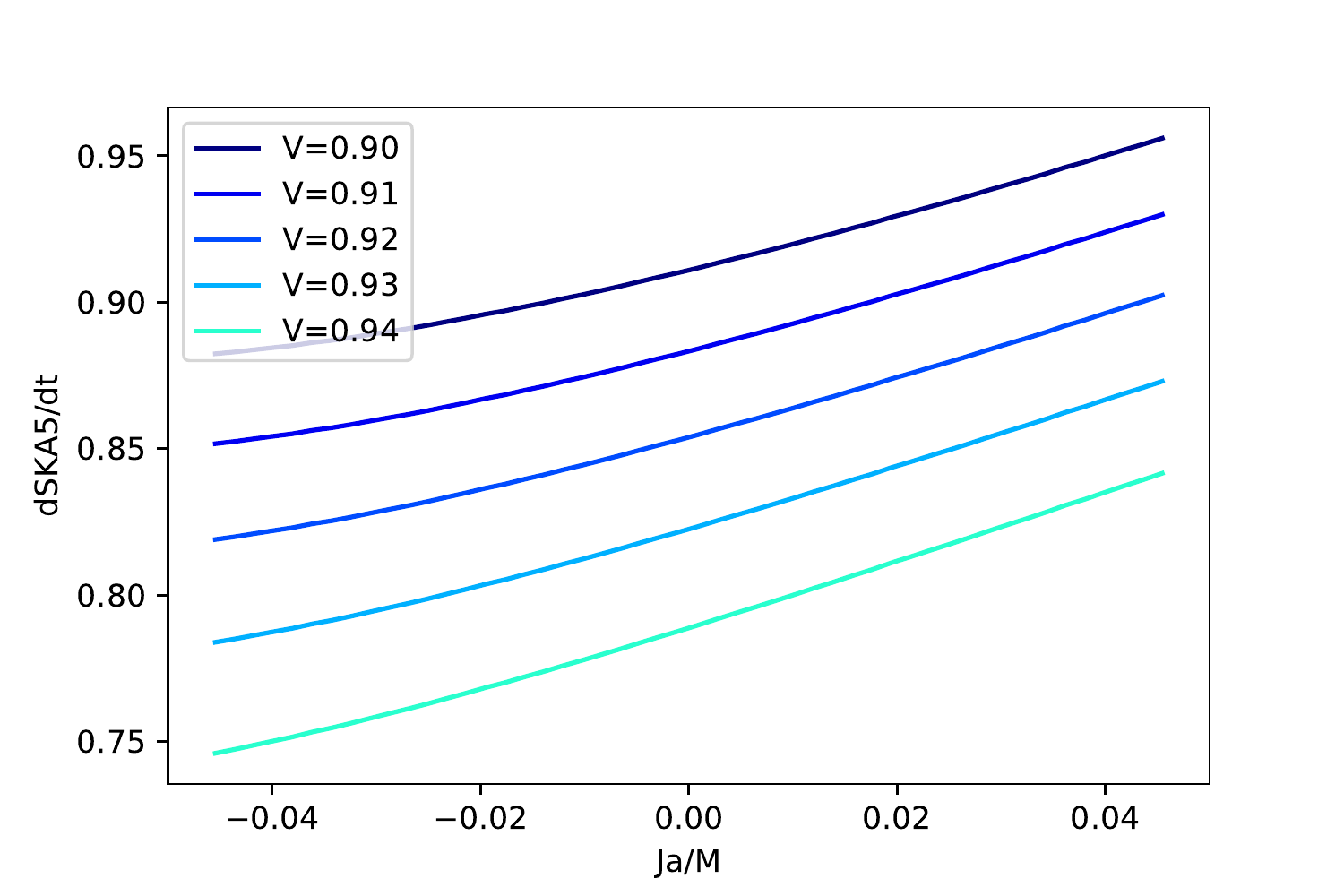}
	\caption{Kerr-AdS$_{4+1}$: Action growth - Black hole $J_a/M$ ($m=5$)}
	\label{fig:KA5aSNGam5}
	\end{minipage}
\hspace{0.01\linewidth}
	\begin{minipage}[t]{0.5\linewidth}
	\includegraphics[width=\linewidth]{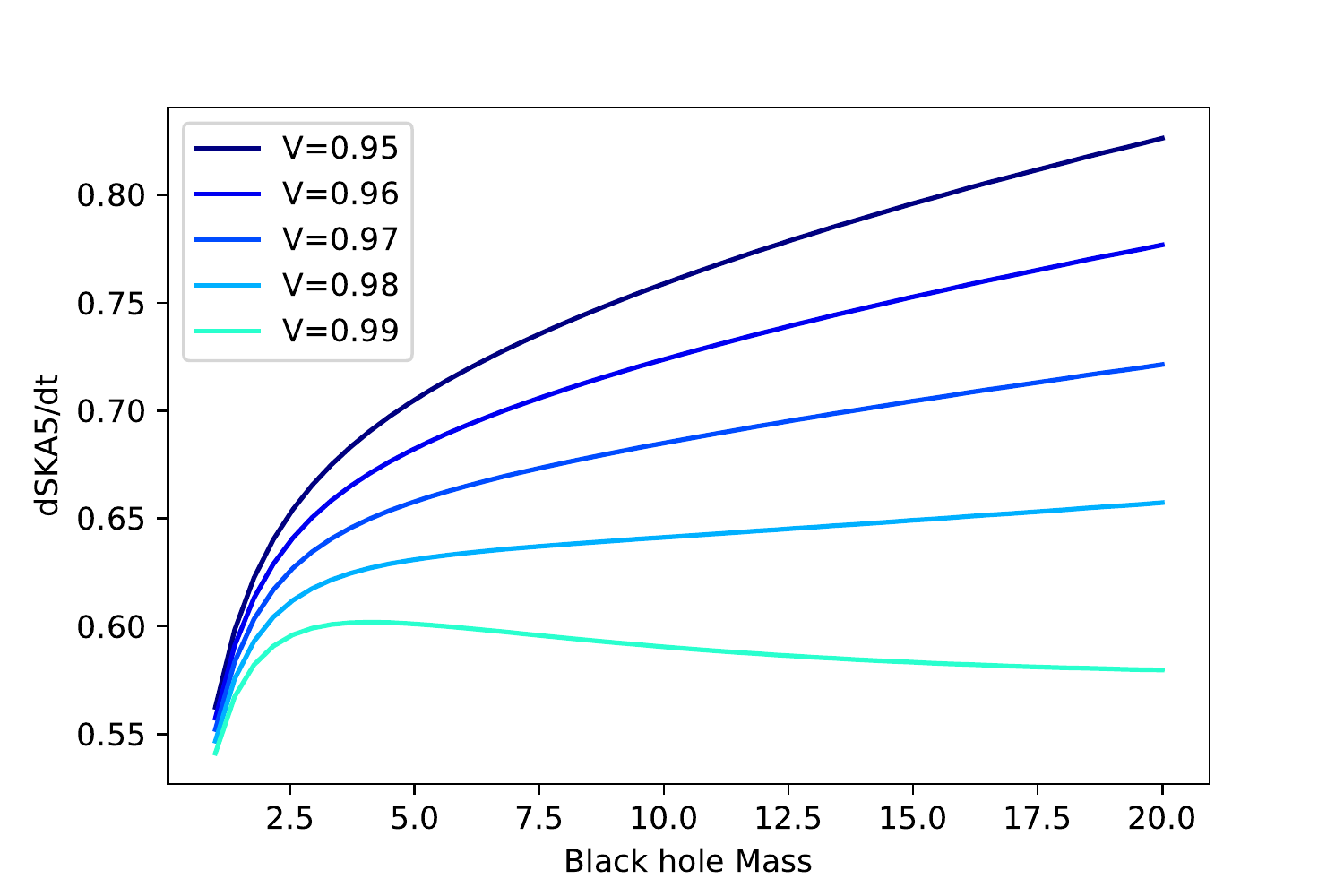}
	\caption{Kerr-AdS$_{4+1}$: Action growth - Black hole mass ($a=0.1$)}
	\label{fig:KA5aSNGMa01}
	\end{minipage}
\end{figure}

The angular momentum per unit mass ($a = J/M$) dependence is shown in figure \ref{fig:KA5aSNGam5}.
Similar to the velocity dependence, a sharp slope tends to disappear.
There is a universal behavior --- the effect of the string is small when the relative velocity is large.

The mass dependence is shown in figure \ref{fig:KA5aSNGMa01}. 
Basically the effect increases according to the black hole mass but this changes to the decreasing function if once the string velocity exceeds the threshold (it is almost the light speed $\approx 0.98c$).


\subsubsection{$b\neq 0$ case}\label{subsec:KA5b}
Next let us consider the $b\neq 0$ case.
In this case the string moves in the axis to the black hole.
The metric becomes
\begin{align}
ds_\text{KA5b}^2
&= -\frac{\Delta_{rb}}{r^2+b^2}dt^2
 + \frac{r^2+b^2}{\Delta_{rb}}dr^2 
 + \frac{r^4(1-b^2)+b^2r^2(1+r^2)}{r^2+b^2}d\phi_1^2,\nonumber\\
&\qquad
\Delta_{rb} = (r^2+b^2)(r^2+1)-2m,\nonumber\\
&= -\Big(r^2+1-\frac{2m}{r^2+b^2}\Big)dt^2
 + \Big(r^2+1-\frac{2m}{r^2+b^2}\Big)^{-1}dr^2 
 + r^2d\phi_1^2.\label{eq:dsKA5b}
\end{align}
Since $\Xi_a = 1$, from eq.\eqref{eq:KA5relMJ} the angular momentum per black hole mass is 
\begin{equation}
\frac{J_b}{M} = \frac23\frac{b}{1+r_{+}^2}.
\end{equation}
We choose the same parametrization as before:
\begin{equation}
\tau = t,\;
r = \sigma,\;
\phi_1 = v\tau + \xi(\sigma).
\end{equation}
Note that the above metric is already the AdS form.
Then one does not need to shift the velocity \eqref{eq:Phishift} to relate the string velocity to the parameter $v$ ($V=v$).
The induced metric is 
\begin{align}
ds_\text{KA5b:ind}^2
&= -f(\sigma)d\tau^2
 + \frac{d\sigma^2}{f(\sigma)}
 + \sigma^2(vd\tau + \xi'd\sigma)^2;\;
f(\sigma) 
:= 1+\sigma^2-\frac{2m}{\sigma^2+b^2},\nonumber\\
&= -(f(\sigma) - v^2\sigma^2)d\tau^2
 + \Big(\frac1{f(\sigma)}+\sigma^2\xi'^2\Big)d\sigma^2
 + 2v\sigma^2\xi'd\tau d\sigma.
\end{align}

\paragraph{EOM and its solution}
This induced metric is the same form as AdS$_{n+1}$ \eqref{eq:dsAdS(n+1)ind} case except that the function $f(\sigma)$ is replaced.
Then, the equation of motion is now obtained only by replacing with the old $f(\sigma)$ with the new one,
\begin{equation}
\xi'(\sigma) 
= \frac{c_\xi}{\sigma^2f(\sigma)}\frac{\mathcal L_\text{KA5b}}{T_\text{s}}
= \frac{c_\xi}{\sigma^2 f(\sigma)} \sqrt\frac{\sigma^2f(\sigma) - v^2\sigma^4}{\sigma^2f(\sigma) - c_\xi^2},\;
f(\sigma) = 1+\sigma^2-\frac{2m}{\sigma^2+b^2}.
\end{equation}
The reality condition in the square root should be imposed. 
The numerator is 
\begin{align}
& 0 = 1 + (1-v^2)\sigma^2 - 2m/(\sigma^2+b^2)\nonumber\\
&\Rightarrow 
(1-v^2)\sigma^4 + (1+(1-v^2)b^2)\sigma^2 + b^2-2m = 0.
\end{align}
Since $D = (1+(1-v^2)b^2)^2 + 4(1-v^2)(2m-b^2) = (1 - (1-v^2)b^2)^2 + 8m(1-v^2) > 0$, this equation certainly has real solutions.
$\sigma_\text{H}$ denotes a positive one of them: 
\begin{equation}
\sigma_\text{H}^2
= -\frac{1+(1-v^2)b^2}{2}+\frac12\sqrt{(1-(1-v^2)b^2)^2+8m(1-v^2)}.
\end{equation}
From the condition for the denominator, the constant $c_\xi$ is determined as $c_\xi^2 = \sigma_\text{H}^2f(\sigma_\text{H})$.
The Lagrangian becomes
\begin{align}
\frac{\mathcal L_\text{KA5b}}{T_\text{s}} 
&= \sqrt\frac{\sigma^2f(\sigma) - v^2\sigma^4}{\sigma^2f(\sigma) - \sigma_\text{H}^2f(\sigma_\text{H})}\nonumber\\
&= \sigma\sqrt\frac{(1-v^2)(\sigma^2+\sigma_\text{H}^2+b^2) + 1}{(\sigma^2+b^2)(\sigma^2+ \sigma_\text{H}^2+1) - 2mb^2/(\sigma_\text{H}^2+b^2)}.
\end{align}
The horizon is determined by
\begin{equation}
\Delta_{rb}(r) = 0 \qquad
\therefore
r_\text{h} = \Big(-\frac{1+b^2}{2}+\frac12\sqrt{(1-b^2)^2+8m}\Big)^{1/2}.
\end{equation}

\begin{figure}[h]
	\begin{minipage}[t]{0.5\linewidth}
	\includegraphics[width=\linewidth]{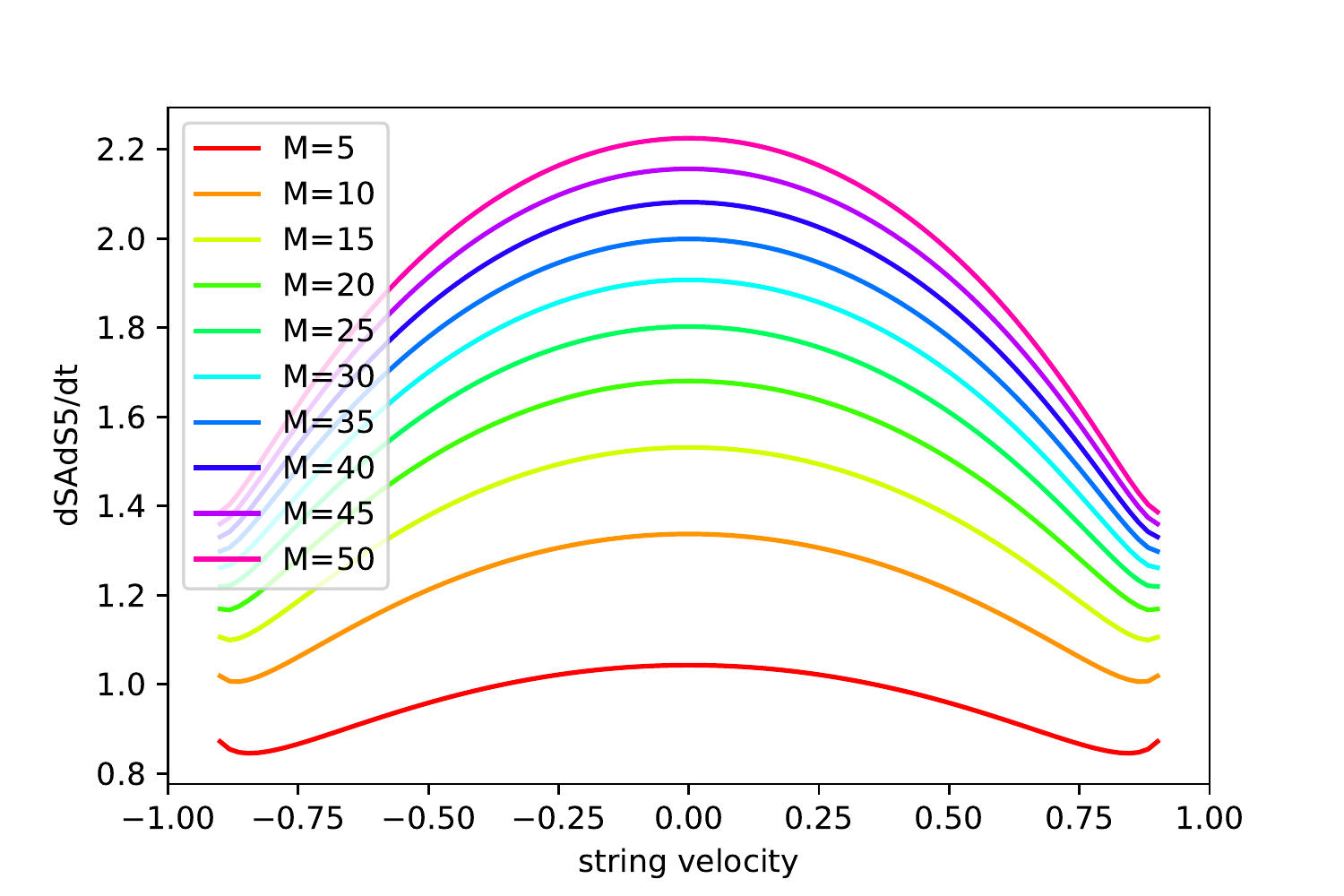}
	\caption{Kerr-AdS$_{4+1}$: Action growth - velocity ($b = 0.1$)}
	\label{fig:KA5ActVb01}
	\end{minipage}
\hspace{0.01\linewidth}
	\begin{minipage}[t]{0.5\linewidth}
	\includegraphics[width=\linewidth]{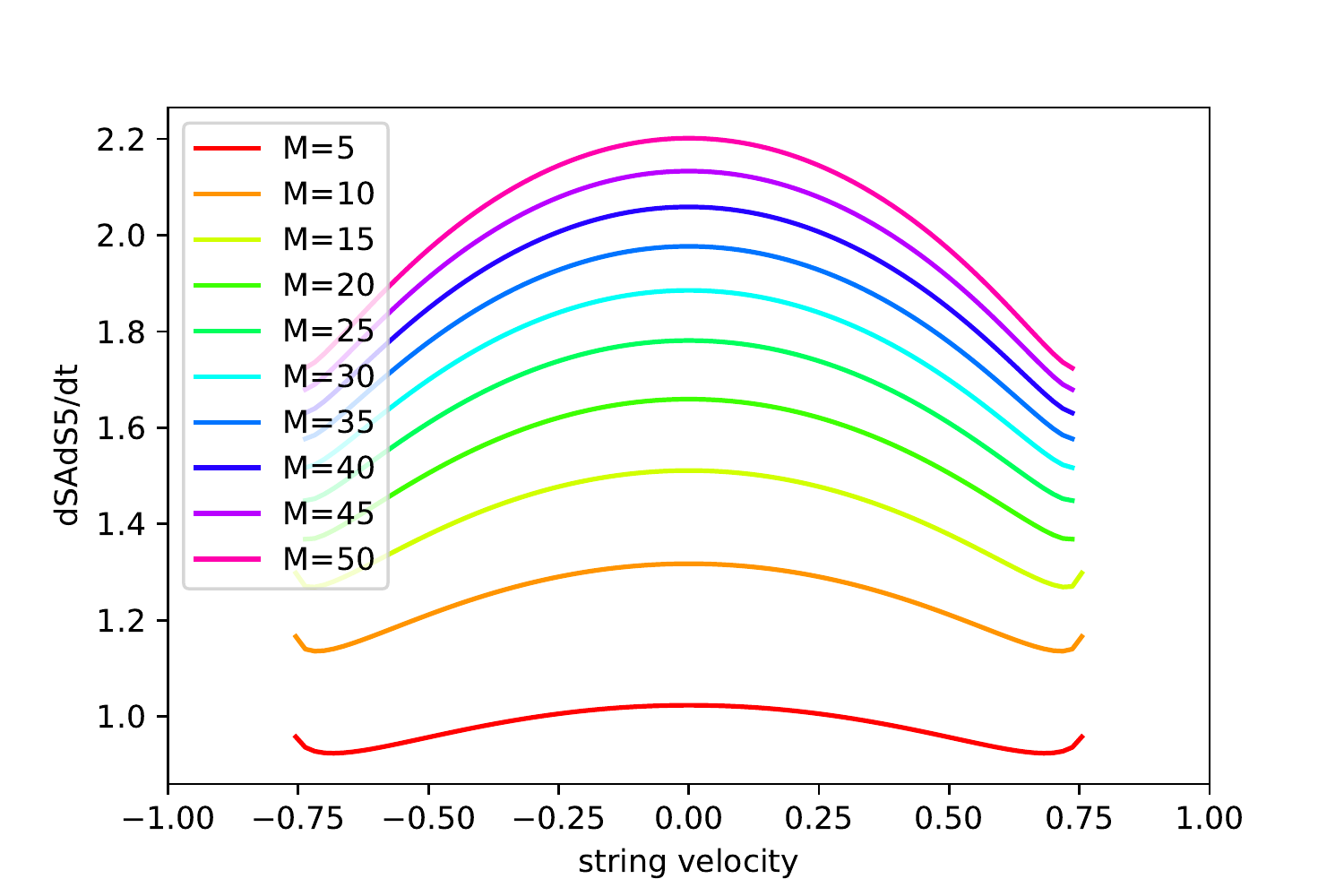}
	\caption{Kerr-AdS$_{4+1}$: Action growth - velocity ($b =  -0.2$)}
	\label{fig:KA5ActVbn02}
	\end{minipage}
\end{figure}

\begin{figure}[h]
	\begin{minipage}[t]{0.5\linewidth}
	\includegraphics[width=\linewidth]{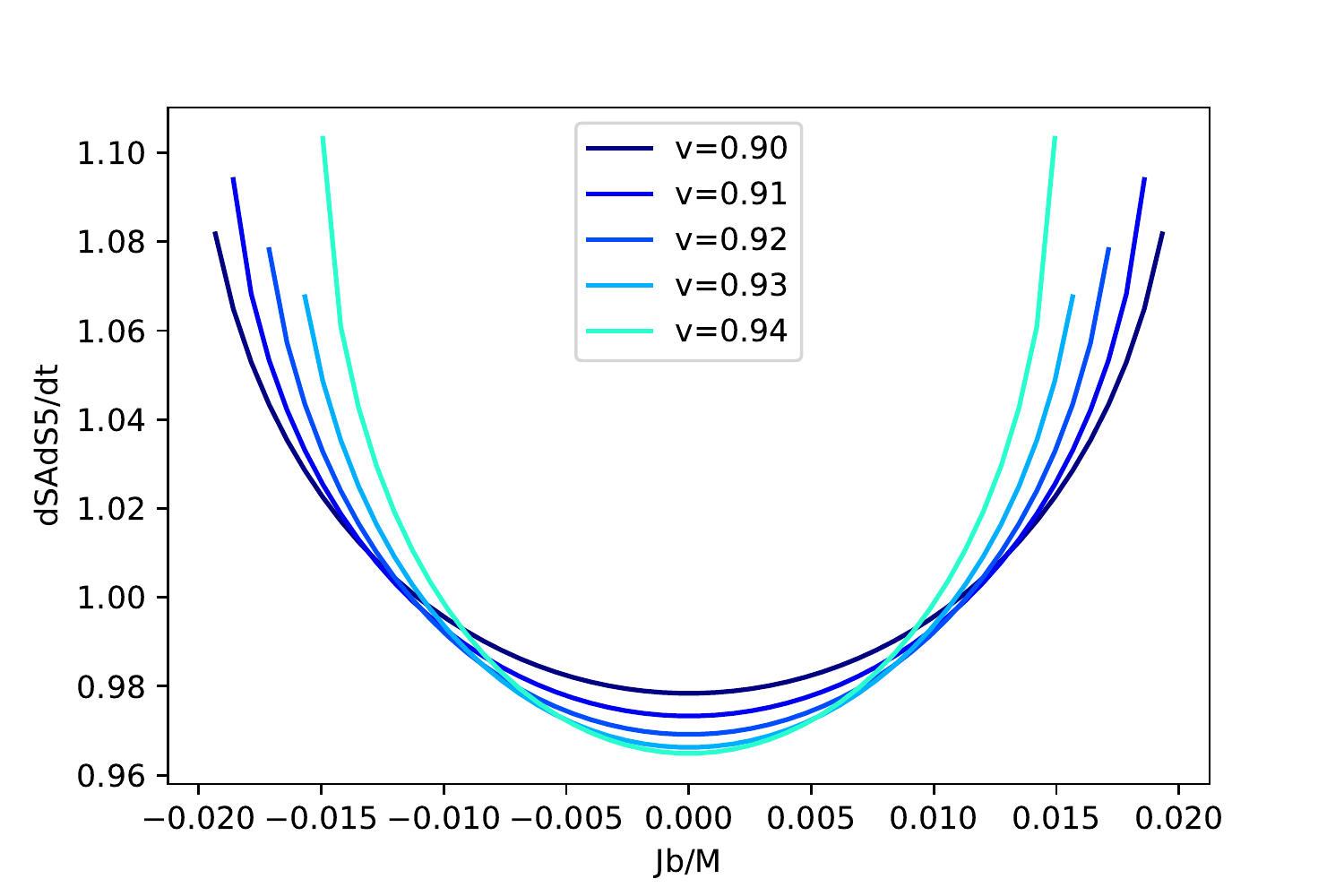}
	\caption{Kerr-AdS$_{4+1}$: Action growth - velocity ($b = 0.1$)}
	\label{figs:KA5ActJbMb01}
	\end{minipage}
\hspace{0.01\linewidth}
	\begin{minipage}[t]{0.5\linewidth}
	\includegraphics[width=\linewidth]{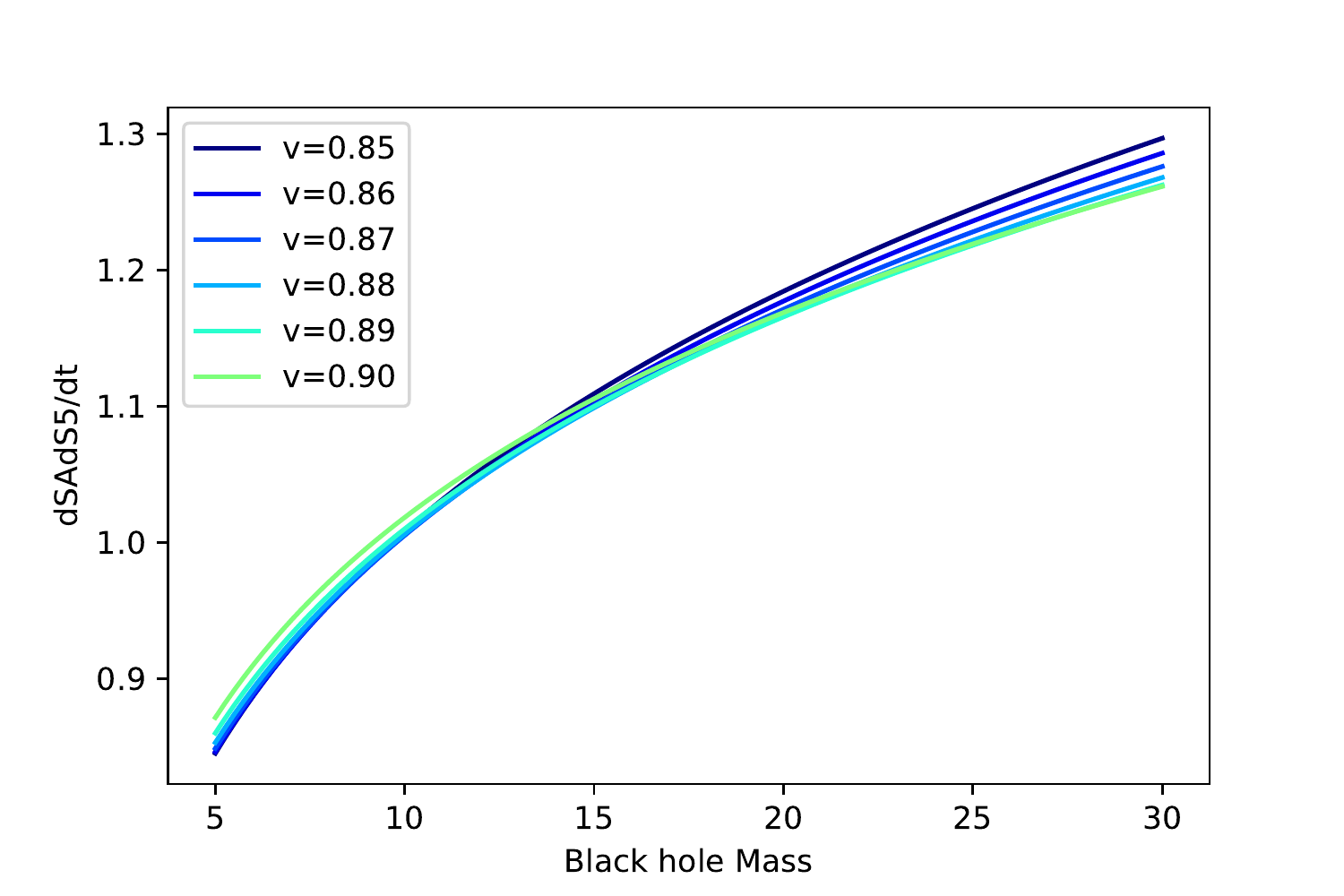}
	\caption{Kerr-AdS$_{4+1}$: Action growth - Black hole mass ($b = 0.1$)}
	\label{fig:KA5ActMb01}
	\end{minipage}
\end{figure}

\paragraph{Action}
Then the development of the NG action obtained by integrating over the WDW patch is 
\begin{align}
\frac{dS_\text{NG}}{dt}
&= \int_0^{r_\text{h}}d\sigma
  \sigma\sqrt\frac{(1-v^2)(\sigma^2+\sigma_\text{H}^2+b^2) + 1}{(\sigma^2+b^2)(\sigma^2+ \sigma_\text{H}^2+1) - 2mb^2/(\sigma_\text{H}^2+b^2)}\nonumber\\
&= \frac12\int_0^{r_\text{h}^2}d\sigma^2
  \sqrt\frac{(1-v^2)(\sigma^2+\sigma_\text{H}^2+b^2) + 1}{(\sigma^2+b^2)(\sigma^2+ \sigma_\text{H}^2+1) - 2mb^2/(\sigma_\text{H}^2+b^2)}.
\end{align}
The figures \ref{fig:KA5ActVb01}, \ref{fig:KA5ActVbn02}, \ref{figs:KA5ActJbMb01} and \ref{fig:KA5ActMb01} show the result of the numerical calculation.

The velocity dependence is shown in figure \ref{fig:KA5ActVb01} and figure \ref{fig:KA5ActVbn02}.
As usual the effect to complexity takes extremum when string is stationary.
There is an abnormal behavior in the vicinity of the light speed.
It tends to increase a bit in the maximum of the velocity.
Because of the reality condition the velocity can not reach to the light speed.
This restricted region becomes narrow according to in creasing the absolute value of $b$.
This is a new phenomena we found.

The dependence on the angular momentum is shown in figure \ref{figs:KA5ActJbMb01}.
Since the string rotates in a different axis, the relative velocity never becomes zero.
Then this behaves very differently from the previous ones.

Figure \ref{fig:KA5ActMb01} shows the mass dependence.
The difference between different velocities ceases and there is no extremum point in this case.
While in the small mass region fast strings gives the large effect, in the large mass region the slower strings the larger effect as usual.


\section{Discussion}\label{sec:Discussion}
\subsection{Summary}
We have seen the effect of the probe string in BTZ, AdS$_{3+1}$, AdS$_{4+1}$, AdS$_{5+1}$ and Kerr-AdS black hole spacetime.
The previous work \cite{Nagasaki:2017kqe} revealed the effect of the probe string in different masses and string velocities.
We could confirm this result and that is a universal behavior in more broad type of black holes.
More specifically, complexity shows different behavior according to the string velocity, black hole mass and the spacetime dimension.
Let us summarize these dependence and its physical interpretation here.

\paragraph{Velocity dependence}
A stationary string gives the maximal complexity growth. 
Complexity decreases as the probe string moves faster.
It seems to contradict the physical intuition because complexity measures how difficult to create the target state from the initial state which is usually stationary.
The same phenomenon was found also in the previous work \cite{Nagasaki:2017kqe}. 
Then we can conclude that this is a universal property of complexity. 

The position of the maximum is shifted in the rotating black hole.
This is thought to be derived from the relative velocity between the string velocity and the black hole angular momentum.  
That is, the effect to complexity is larger when the relative velocity is smaller.
The maximum value also decreases as the maximum point moves from the center by this shift.
Near the light speed there is also an interesting phenomena in the mass dependence as stated below.

Let us note that the universal property of complexity stated above does not stem from the time delation of relativistic phenomena.
Figures \ref{fig:BTZActionvelocityJM09}, \ref{fig:BTZActionvelocityJM02}, \ref{fig:KA4Actvelocitya01} and \ref{fig:KA4Actvelocityan02} show the peak position shifts according the relative velocity between the black hole and the probe string.
Since we calculated the NG action on the rest flame, the peak locates at $v=0$ if this behavior stems from the time dilation.
We can also see that if this behavior derives from the Lorentz factor $\sqrt{1-v^2}$, it does not behave linearly as BTZ cases (figure \ref{fig:BTZActionvelocityJM09} and figure \ref{fig:BTZActionvelocityJM02}).

\paragraph{Mass dependence}
Complexity basically tends to increase according the mass.
This can be thought that this is because complexity defines how complex of the physical system.

A remarkable phenomenon occurs in the vicinity of the light speed.
In the lower dimension, AdS$_{3+1}$ and the AdS$_{4+1}$, the dependence on black hole mass has the maximum point for near light speed strings.
That maximum disappears for higher dimension as shown in the AdS$_{5+1}$ dimensional case \ref{fig:AdS51Actionmass}. 

In lower dimension, AdS$_{n+1\leq 5}$ the mass dependence can be a decreasing function of mass for a near light speed string.

\paragraph{Dimensionality dependence}
As the spacetime dimension becomes higher, the peak of the dependence on the string velocity becomes smooth.
Especially, in three dimensional case, the velocity dependence in BTZ black holes forms a broken line.
As the dimensionality becomes higher this slope tend to become gentle.

We can conclude that the effect of the probe string becomes insensitive in higher dimensions.
It can be intuitively explained as follows.
Although the Nambu-Goto action is proportional to the two dimensional world volume in whole spacetime, we restrict the motion of the string in $S^1$ subspace of a specific plane.
In order to remove this restriction, we investigate in section \ref{subsec:KA5b} the case where string moves around a different axis to the black hole angular momentum.
As expected a new phenomena was found.
That is, the dependence on the string velocity does not decrease near the light speed (see figs \ref{fig:KA5ActVb01} and \ref{fig:KA5ActVbn02}).
Furthermore, the difference of the dependence on mass in various string velocities disappears in this case (see fig \ref{fig:KA5ActMb01}).

\begin{figure}[h]
	\begin{minipage}[t]{0.5\linewidth}
	\includegraphics[width=\linewidth]{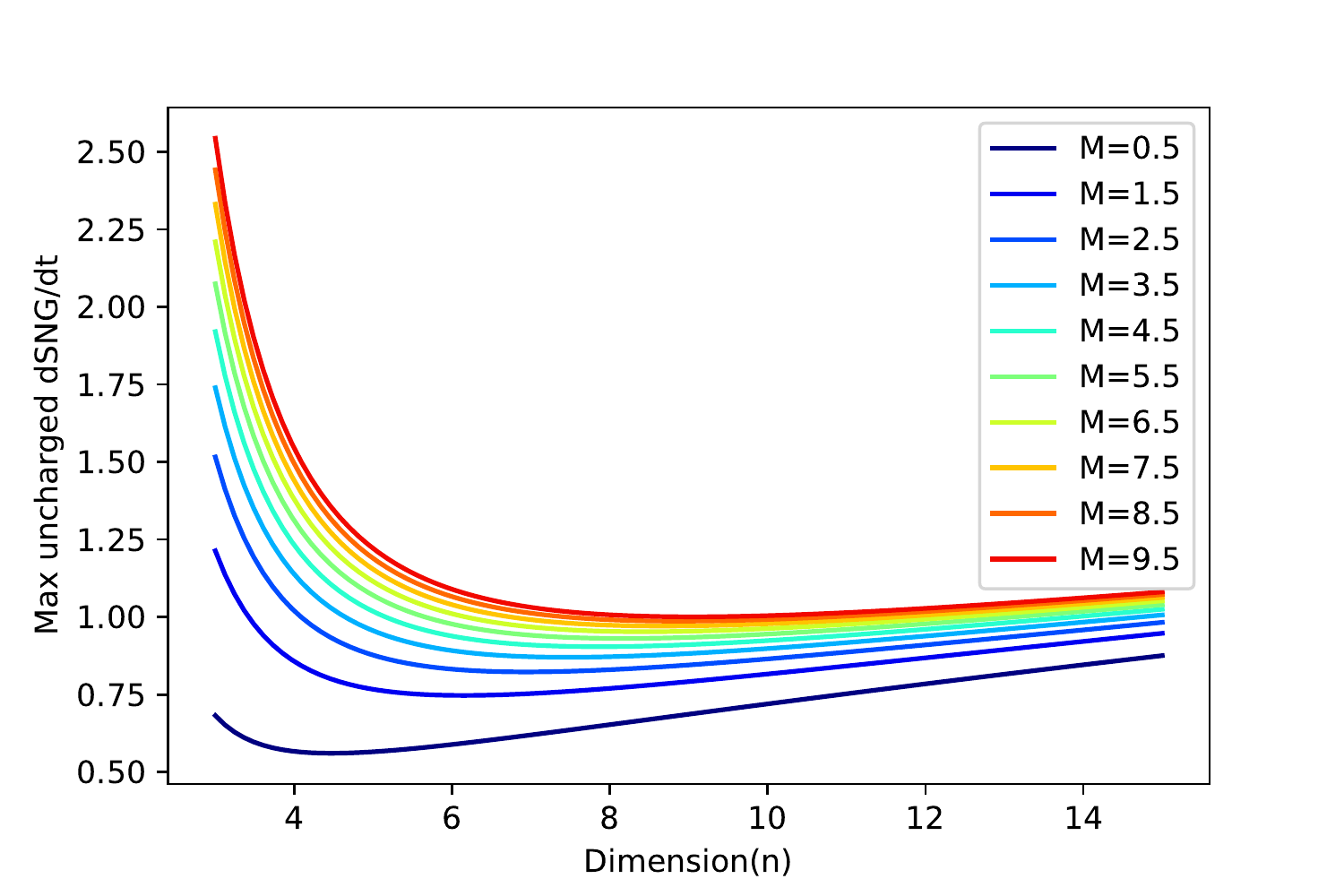}
	\caption{Maximum of the action growth in diverse dimensions (uncharged case)}
	\label{fig:dSdtMaxAdS(n+1)}
	\end{minipage}
\hspace{0.01\linewidth}
	\begin{minipage}[t]{0.5\linewidth}
	\includegraphics[width=\linewidth]{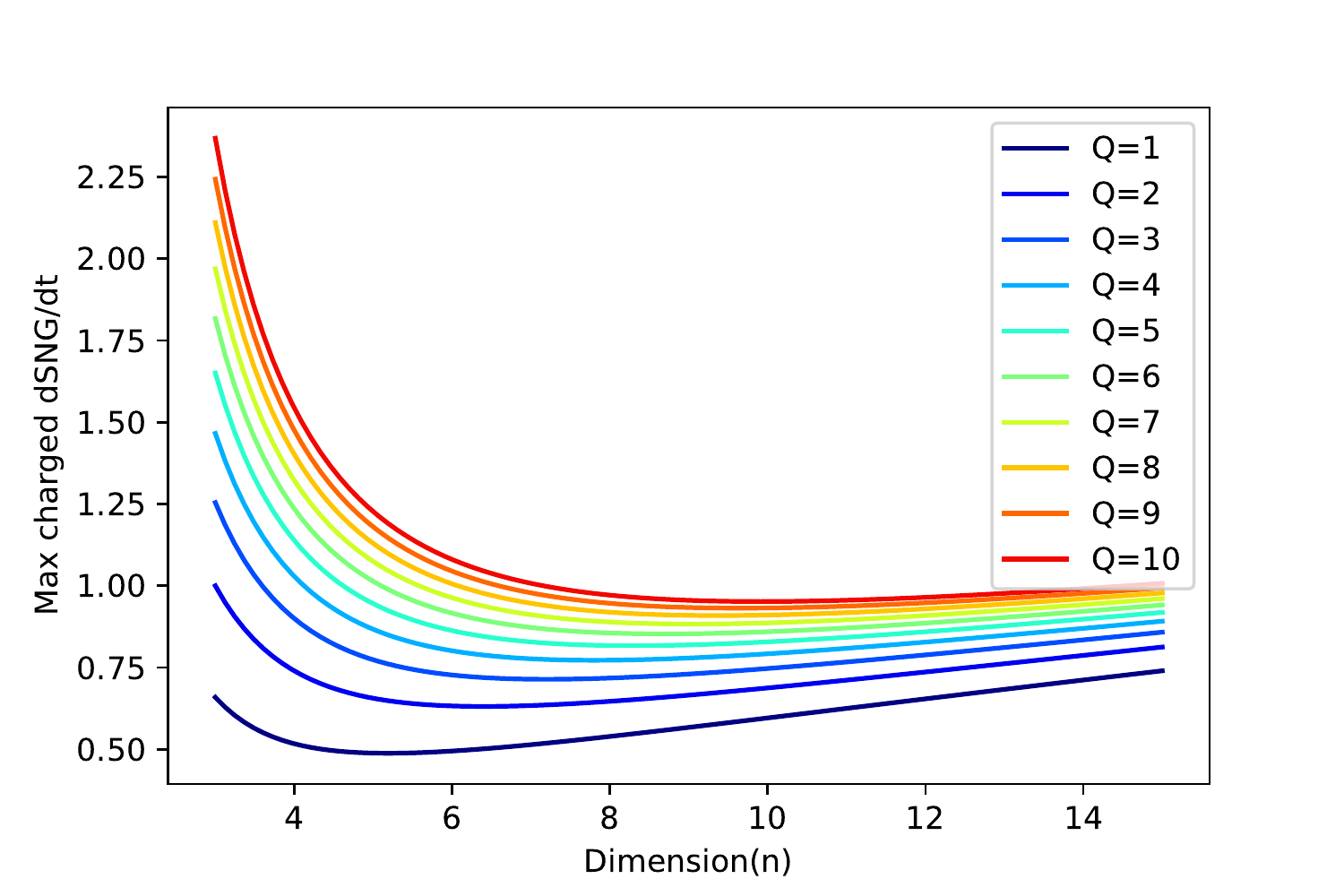}
	\caption{Maximum of the action growth in diverse dimensions (charged case)}
	\label{fig:dSdtMaxQAdS(n+1)}
	\end{minipage}
\end{figure}

\paragraph{Maximum value}
According to the results in sec.\ref{sec:AdSBH} (figures \ref{fig:AdS31ActionVelocity}, \ref{fig:AdS41Actionvelocity} and \ref{fig:AdS51Actionvelocity}), 
the plots of the velocity dependence not only becomes smoother, but its maximum value also looks to decrease.
Let us confirm whether this behavior is universal.
We focus on AdS$_{n+1}$ black holes.
We know already that the effect of the string is maximum when the string velocity is zero.
For $v=0$, the Lagrangian \eqref{eq:AdSNGaction(n+1)} is unity.
The integral depends only on the horizon:
\begin{equation}\label{eq:SNGrh}
\frac{\mathcal L_\text{AdS(n+1)}}{T_\text{s}}
= r_\text{h}.
\end{equation}
The horizon $r_\text{h}$ is determined by (see the metric function \eqref{eq:AdSmetricfunc(n+1)})
\begin{equation}
0 = f(r)
= 1 - \frac{16\pi}{(n-1)\Omega_{n-1}}\frac{M}{r^{n-2}} + r^2
\Rightarrow
r^n + r^{n-2} - \frac{8\pi^{-n/2+1}\Gamma(n/2)}{n-1}M = 0.
\end{equation}
The maximum value of the NG action in diverse dimensions is plotted in figure \ref{fig:dSdtMaxAdS(n+1)}.

There is a difficulty for the charged case, as I explain later below of eq.\eqref{eq:AppsigmaH}.
But only the maximum value can be obtained in the same way as the uncharged case.
We focus on the extremal black holes.
The Lagrangian is again unity and the action is equal to the horizon.
This horizon is determined in the same way by
\begin{equation}\label{eq:horizoneqQ}
r^n + r^{n-2} 
- \frac{4\pi^{-n/2+1}\Gamma(n/2)}{n-1}\Big(2M - \frac{Q^2}{r}\Big) = 0.
\end{equation}
Given a charge, the extremal mass is determined by setting the minimum value of the left hand side \eqref{eq:horizoneqQ} is zero.
The dimensionality dependence of the maximum value of the NG action for extremal black holes is plotted in figure \ref{fig:dSdtMaxQAdS(n+1)}.
By this plots we see that the maximum value decreases at lower dimensions.
Furthermore, the minimum point approaches ten-dimension for sufficiently large charges.

\subsection{Future direction}
In this work we considered the effect of the probe string. 
That corresponds to the introduction of a kind of nonlocal operator --- a Wilson loop.
We first expect the generalization of the dimension.
Several higher dimensional local operators can be added.
Especially, the co-dimension local operator, interface, is an interesting object.
This local operators realized in a system consisting of two kind branes --- D3/D5.
Since complexity is known to have a nonlocal property, it must be useful to use such kind of operators to study the property of complexity.
Some interesting properties of the nonlocal operators in BTZ black holes are already found in \cite{Ageev:2014nva} and complexity growth of defect theory is studied in \cite{Ovgun:2018jbm}.
One suggest is to study the effect of these nonlocal operators in the diverse kinds of black holes.

In \eqref{eq:AdSmetric(n+1)} we restrict the case for uncharged AdS black holes.
The growth of the Einstein-Hilbert action for charged case is studied in \cite{Cai:2017sjv}.
I would like to study the nonlocal operators in these kinds of black holes.
The adding of the charge is an important future work since it is related to check whether the complexity growth satisfies the Lloyd bound \cite{Brown:2015lvg, 2000Natur4061047L}.
But in this case a difficulty occurs.
The metric function in this case is 
\begin{subequations}
\begin{align}
ds_\text{AdS$_{n+1}$}^2
&= -f(r)dt^2 + \frac{dr^2}{f(r)} + r^2d\Omega_{n-1},\\
f(r) 
&= 1 - \frac{8\pi}{(n-1)\Omega_{n-1}}\Big(\frac{2GM}{r^{n-2}}-\frac{GQ^2}{r^{n-1}}\Big)
  + \frac{r^2}{\ell_\text{AdS}^2}
= 1 - \frac{r_\text{m}^{n-2}}{r^{n-2}} + \frac{r^2}{\ell_\text{AdS}^2},\\
& 
r_\text{m}^{n-2} := \frac{16\pi GM}{(n-1)\Omega_{n-1}},\;\;
r_\text{q}^{n-1} := \frac{8\pi GQ^2}{(n-1)\Omega_{n-1}},
\end{align}
\end{subequations}
The condition for the numerator \eqref{eq:realcondnum} is changed by adding a new term as
\begin{align}\label{eq:AppsigmaH}
(1-v^2)\sigma^n + \sigma^{n-2} - r_\text{m}^{n-2} + r_\text{q}^{n-1}/\sigma = 0.
\end{align}
The left hand side of this equation is no longer monotonically increasing. 
If this function has real solutions there are two solutions at least (including multiple).
So the same procedure can not be used since we need to determine the constant $c_\xi$ in the denominator \eqref{eq:solxindim} using the unique solution of the above equation.

\section*{Acknowledgement}
I would like to thank Satoshi Yamaguchi, UESTC and KEK members and people discussing at 73th Physical Society of Japan Annual Meeting for helping my research.


\providecommand{\href}[2]{#2}\begingroup\raggedright\endgroup

\end{document}